\documentclass[%
 reprint,
 superscriptaddress,
 amsmath,amssymb,
 aip,
 jcp,
 floatfix,
]{revtex4-2}
\usepackage{graphicx}
\usepackage{dcolumn}
\usepackage{bm}

\usepackage{amsmath}
\usepackage{amssymb}
\usepackage{hyperref}
\usepackage{cleveref}
\usepackage{graphicx}
\usepackage{color}
  
\usepackage[normalem]{ulem}

\begin{document}

\title{Relevance of Aggregate Anisotropy in Sheared Suspensions of Carbon Black}

\author{Victor T\"anzel}
\affiliation{Statistical Physics of Soft Matter and Complex Systems, Institute of Physics, University of Freiburg, Freiburg, Germany}%
\affiliation{Cluster of Excellence \textit{liv}MatS @ FIT – Freiburg Center for Interactive Materials and Bioinspired Technologies, University of Freiburg, Georges-Köhler-Allee 105, D-79110 Freiburg, Germany}%

\author{Fabian Coupette}
\affiliation{School of Mathematics, University of Leeds, Leeds, United Kingdom}%

\author{Marisol Ripoll}
\affiliation{Theoretical Physics of Living Matter, Institute for Advanced Simulation, Forschungszentrum J\"ulich, 52428 J\"ulich, Germany}%

\author{Tanja Schilling}
\email[]{tanja.schilling@physik.uni-freiburg.de}
\affiliation{Statistical Physics of Soft Matter and Complex Systems, Institute of Physics, University of Freiburg, Freiburg, Germany}%

\date{\today}

\begin{abstract}
    Carbon Black is a filler frequently used in conductive suspensions or nanocomposites, in which it forms networks supporting electric conductivity.
    Although Carbon Black aggregates originate from a presumably isotropic aggregation process, the resulting particles are inherently anisotropic. Therefore, they can be expected to interact with shear flow, which significantly influences material properties.
    In this study, we investigate sheared suspensions of Carbon Black aggregates to elucidate the impact of aggregate anisotropy on the rheological properties. 
    We aim at
    concentrations below and above the conductivity percolation threshold and
    comprehensively characterize particle behavior under flow conditions.
    Aggregates assembled by a diffusion-limited aggregation process are simulated with Langevin dynamics in simple shear flow.
    The simulations reveal a clear alignment of the aggregates’ long axis with the flow direction, an increase in tumbling frequency with higher shear rates, and a shear-thinning response. This behavior closely parallels that of rod-like particles and underlines the significance of the anisotropic nature of Carbon Black aggregates.
    These findings will facilitate the optimization of nanocomposite precursor processing and the tailoring of Carbon Black–based conductive suspensions.
\end{abstract}

\maketitle

\section{Introduction}

Suspensions of electrically conductive particles dispersed in insulating fluids are central to a variety of technologies, from printable inks \cite{loffredo2009ink,cui2016printed,camargo2021development,cummins2012inkjet}, flowable sensors \cite{camargo2021development,schmidt2024electrofluids} to flow electrodes \cite{presser2012electrochemical,campos2013investigation,zhang2021flow,alfonso2021highly} and liquid precursors of conductive nanocomposites \cite{wonisch2011comprehensive,choi2019high,li2019review,kanoun2021review}. 
These applications exploit that the embedded colloidal particles form transient networks which span the fluid, rendering the suspension conductive whilst maintaining the mechanical and rheological properties of a liquid or soft solid. 
Shear flows frequently occur for these applications and can be expected to drive the dynamical behavior of suspended elongated particles and influence material properties like conductivity, mechanical response, or electromagnetic shielding \cite{thomassin2013polymer} of the suspensions. 
It is thus imperative to develop a thorough understanding of the filler behavior under shear flow to optimize the material, ultimately facilitating the design and fabrication of flowable electronics.

Many studies have analyzed suspensions featuring conductive particles with a variety of regular and well controlled geometries such as spherical metal particles, cylindrical Carbon nanotubes, or planar graphene layers \cite{camargo2021development,cummins2012inkjet,tan2019metallic}.
In this work, we scrutinize the flow properties of Carbon Black aggregates, i.e., industrially produced soot particles \cite{watson2001carbon}.
Carbon Black aggregates consist of globules of turbostratic graphene layers, so-called primary particles, rigidly fused together during the incomplete combustion of fuel in industrial furnaces at very high temperatures. 
As a consequence, the arrangement of primary particles is highly disordered and self-similar on intermediate length scales.
The process of aggregation itself is difficult to model dynamically due to the abundance of uncontrolled parameters in the combustion process. 
Instead, the resulting aggregates are analyzed statistically and the relevant aggregation mechanism is inferred from the aggregate characteristics.
Although the most frequently deployed aggregation schemes such as diffusion-limited, reaction-limited, or cluster-cluster aggregation are isotropic, small aggregates in particular are far from spherical simply due to random fluctuations in the generation process.
Therefore, aggregates may exhibit individual and complex flow behavior which could have substantial impact on the macroscopic material properties.
At the same time, Carbon Black is cheap in production compared to more elaborate carbon varieties such as carbon nanotubes while still becoming conductive at relatively low volume fractions \cite{kluppel2014role,coupette2021percolation}.
Accordingly, it is very attractive for industrial applications ranging from mechanical reinforcement in rubbers to providing electrical conductivity in printable inks.
All these applications subject Carbon Black suspensions to flow, either during the fabrication of precursor materials or in the product itself.
In this article, we therefore examine the impact of the fractal microscopic morphology and effectively elongated shape of Carbon Black aggregates on the rheological properties of the corresponding suspension.

Previous work has explored the dynamics of isolated aggregates, commenting on tumbling behavior \cite{fellay2013motion, trofa2020rheology, yu2024motion}, 
fracture of unstable aggregates \cite{asylbekov2021microscale, dizaji2019collision}, 
sedimentation \cite{trofa2020sedimentation},
and viscosity \cite{trofa2020rheology}.
However, to the best of our knowledge, a comprehensive, ensemble-level study of how the anisotropy of fractal aggregates affects particle alignment and rheology at finite concentrations under steady shear is lacking. 
To fill this gap, we generate Carbon Black-like aggregates by diffusion-limited aggregation to reproduce fractal, anisotropic morphologies. 
Using Langevin dynamics simulations, we probe their non-equilibrium steady states in simple shear. 
This approach allows us to quantify orientation statistics and alignment, tumbling dynamics, and viscosity trends for suspensions of interacting aggregates. 
To address the context of conductive suspensions and nanocomposites, we focus on volume fractions near the conductivity percolation threshold of Carbon Black \cite{coupette2021percolation}.

Particles produced by diffusion-limited aggregation embedded in a solvent do not only serve as a model for Carbon Black.
They have for example also been studied in the context of colloidal aggregates in general \cite{poon1997mesoscopic}, thermal conductivity of nanofluids \cite{evans2008effect}, contrast agents for medical imaging \cite{etheridge2014accounting}, ocean sediments \cite{yu2024motion}, and interstellar dust \cite{katyal2014fractal}. 
Consequently, the results we present here apply generally for fractal aggregates across domains.

\section{Methods}

\subsection{Carbon Black model}
Carbon Black has been studied extensively in experiments and the structure of its constituting particles can be controlled during the industrial production process, leading to primary particles which are all roughly of the same size.
Collisions cause them to fuse into rigid aggregates, the second layer in the hierarchical structure of carbon black and the subject of our investigation.
This aggregation process is intrinsically random and leads to mass fractal particles, meaning that the number of primary particles within a given radius measured from the center of mass of the aggregate scales as a power law with exponent $D_f$ called fractal dimension. 
For Carbon Black aggregates, the fractal dimension $D_f$ has been reported  \cite{meakin1983diffusion,kluppel1995fractal} to be between $2.2$ and $2.8$.
Diffusion limited aggregation is one of the simplest and most studied aggregation models which happens to generate aggregates with a fractal dimension of roughly 2.5, making it an excellent model for common Carbon Black varieties \cite{meakin1983diffusion}. 
Aggregates can cluster on larger length scales due to van der Waals forces to form agglomerates, the third and final hierarchical layer. In this paper we focus on the steric ramifications of the fractal morphology, i.e., purely repulsive aggregates, so that agglomerates will not be addressed here.  

\begin{figure}
    \centering
    \includegraphics[width=.8\linewidth]{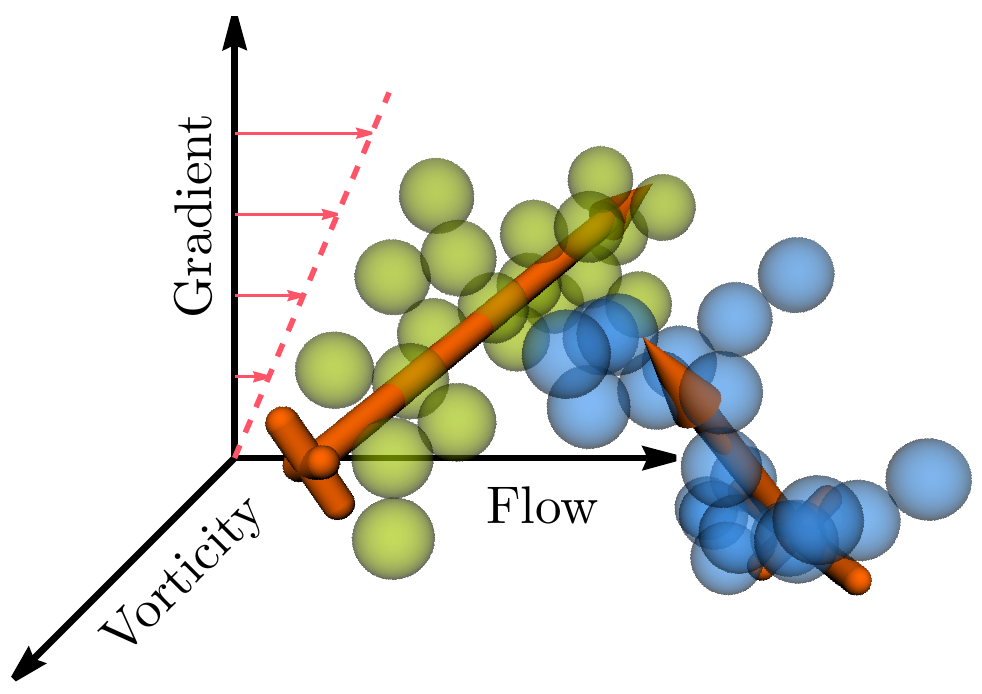}
    \caption{Two interacting example aggregates in a flow geometry. Each aggregate is characterized by its principal axes,  long (arrows), intermediate and short (tubes) axis, respectively.
    The shear flow is described by the flow ($x$), velocity gradient ($y$) and vorticity ($z$) axes.
    }
    \label{fig:aggregates_render}
\end{figure}

Recently, we developed a particle-based model for Carbon Black aggregates \cite{coupette2021percolation,zimmer2024flow}. It is based on diffusion-limited aggregation, in which primary particles diffuse independently and sequentially until they encounter and stick to the growing structure, which originates from a seed particle at the origin. In this first step, we generate many aggregates independently. Next, systems of the resulting aggregates are studied with Monte Carlo simulations, in which they are treated as rigid bodies composed of hard spheres. In \cref{fig:aggregates_render}, we visualize two exemplary aggregates.

To adapt the model to molecular dynamics, the primary particles are treated as purely repulsive Weeks-Chandler-Anderson spheres that interact via the potential
\begin{equation}
    U_\mathrm{WCA}(r) = 
    \begin{cases}
    4\epsilon \left[ \left(\frac{\sigma}{r}\right)^{12} - \left(\frac{\sigma}{r}\right)^6 \right] + \epsilon, & r < 2^{1/6}\,\sigma \\
    0, & r \geq 2^{1/6}\,\sigma\,,
    \end{cases}
\end{equation}
where $\epsilon = 10\,k_\mathrm{B}T$ is chosen for the energy parameter. The interaction is applied only between primary particles belonging to different aggregates. We use the thermal energy $k_\mathrm{B}T$, size parameter $\sigma$ and the mass of primary particles $m$ as units, such that the unit of time can be expressed as $\tau = \sqrt{\sigma^2 m/(k_\mathrm{B}T)}$.
Aggregates are restrained using harmonic bond, angle and dihedral potentials 
of the form $U(r) = k_b (r-r_0)^2$ for the bonds, analogously for the angular potentials.
We infer angles and dihedrals from the bond topology.
The equilibrium distance of the bonds is $r_0=1\,\sigma$, equilibrium angles result from the diffusion-limited aggregation process. As spring constants, we employ $k_b = 450\,k_\mathrm{B}T/\sigma^2$ for bonds, and $k_a = k_d = 100\,k_\mathrm{B}T/\text{rad}^2$ for angles and dihedrals.

We study one system containing $N = 1000$ aggregates, each consisting of $N_p = 20$ primary particles which corresponds to a fractal dimension\cite{coupette2022percolation} of $D_F(N_p)\approx 2.8$. Volume fractions are defined as $\phi = N_pN\pi/(6V)$, the ratio of the sum of primary particle volumes to the system volume $V$.
By changing the box size, we obtain volume fractions of $\phi = 5.96\,\%$, $8.90\,\%$, and $12.23\,\%$, which are close to the percolation threshold
for our system \cite{coupette2021percolation}. In the following, we round the values to the next integer for brevity.

\subsection{Molecular Dynamics}
To study the dynamics of Carbon Black aggregates under shear flow, we use molecular dynamics simulations. The integrator is a Velocity Verlet algorithm with time step $\Delta t = 10^{-4}\,\tau$ as implemented in LAMMPS \cite{thompson2022lammps}. We first set up the studied system of aggregates with the Metropolis Monte Carlo method described in prior work \cite{coupette2021percolation}. Then, we adapted the box size to correspond to the desired volume fraction, and equilibrated the systems again in the molecular dynamics simulations in equilibrium and in shear flow until a steady state was reached.

In order to investigate shear, we utilize Lees-Edwards boundary conditions to generate a linear shear flow profile. This defines the three laboratory frame axes: flow, (velocity) gradient, and vorticity axis, which we associate with the Cartesian axes $x, y, z$, respectively. For the temperature control, we employ a Langevin thermostat with time constant $\tau_L = 0.1\,\tau$. This introduces a viscous drag force $-m v /\tau_L$ acting on primary particles, along with stochastic forces. The magnitude of these kicks is $\sqrt{k_\mathrm{B}T m / (\Delta t \tau_L)}$, scaled by uniformly distributed random numbers in the interval $[0,1]$ \cite{brunger1984stochastic,dunweg1991brownian}. We enforce a vanishing total random force to suppress drift. To account for the shear flow, the velocity $v$ is relative to the expected linear shear profile. Measurements were carried out over $1.6 \cdot 10^4\,\tau$ for the alignment and viscosity study as well as over $10^4\,\tau$ for the tumbling analysis.

\section{Results}

\subsection*{Aggregate shape \& diffusion}
\begin{figure}
    \centering
    \includegraphics[]{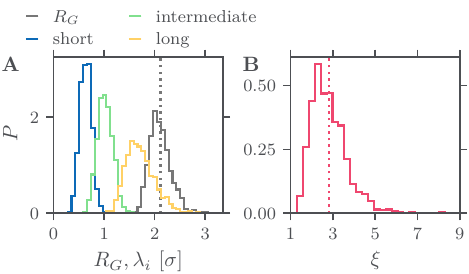}
    \caption{Shape characterization of the aggregates via the probability distributions of their radius of gyration $R_G$ (A), axis lengths $\lambda_i$ (A), and aspect ratio $\xi$ (B). The dotted vertical lines mark the mean radius of gyration $\left< R_G\right>_N \approx 2.11\,\sigma$ and mean anisotropy $\left< \xi \right>_N \approx 2.81$. The aggregates exhibit a range of shapes and a clear deviation from the spherical limit.}
    \label{fig:distribution_Rg_aniso}
\end{figure}

 In the simulations, we use one ensemble of aggregates at varying volume fractions. We begin its characterization with the shape of the aggregates and define the gyration tensor for each aggregate as
\begin{equation}
    G_{\alpha \beta} = \frac{1}{N_p} \sum_i^{N_p} (r_{i,\alpha}-r_{\mathrm{cm},\alpha})(r_{i,\beta}-r_{\mathrm{cm},\beta})\;,
\end{equation}
where $N_p=20$ is the number of primary particles, $\mathbf{r}_i$ is the position of primary particle $i$, $\mathbf{r}_\mathrm{cm}$ is the aggregate center of mass, and $\alpha,\beta \in \{x,y,z\}$. The gyration tensor is also accessible in experiments \cite{schroeder2005dynamics}. Diagonalization yields three orthonormal aggregate axes, i.e., the eigenvectors of the gyration tensor which, in descending order of associated eigenvalue $\lambda_i^2$, correspond to the long, intermediate, and short axis, respectively, of the aggregate. Fig.~\ref{fig:aggregates_render} illustrates these principal axes for two individual aggregates.
To describe the characteristic size of an aggregate, we calculate its radius of gyration
\begin{equation}
   R_G = \sqrt{\lambda_1^2 + \lambda_2^2 + \lambda_3^2}\;,
\end{equation}
which is, for example, the relevant measure in the interpretation of orientation-averaging experimental techniques such as scattering. 
As \cref{fig:distribution_Rg_aniso} shows, the aggregates have a mean radius of gyration of $\left< R_G \right>_N \approx 2.11\,\sigma$, which defines the characteristic size of the ensemble. To quantify the extent of the three aggregate axes, we present distributions of the roots $\lambda_i$ of the eigenvalues of the gyration tensor. Even though their distributions have some overlap, the axes are likely to be distinct from each other for individual aggregates. 

Using the lengths of the long and short aggregate axes $\lambda_1, \lambda_3$, we define an effective aspect ratio
\begin{equation}
    \xi = \frac{\lambda_1}{\lambda_3}\;.
\end{equation}
In the investigated set of aggregates, this parameter is on average $\left< \xi\right>_N \approx 2.81$, but even exceeds $\xi = 4$ for $8.4\,\%$ of the aggregates. Overall, the aggregates are distinctly nonspherical.

For direct comparisons of shear flow simulations, whether between different computational approaches or experimental measurements, it is often not useful to characterize the shear flow strength by the shear rate $\dot\gamma$. The reason for this are different diffusion behaviors and associated time scales. To provide a dimensionless measure, we employ the P\'eclet number
\begin{equation}\label{eq:peclet}
    Pe = \frac{\dot\gamma \left<R_G\right>_N^2}{D}\;.
\end{equation}
It quantifies the ratio of advective and diffusive transport rates, thereby weighing the influence of the imposed shear flow. In order to provide an accurate numerical value of $D$, we measure the mean squared displacement of the centers of mass of aggregates and determine the diffusion constants $D= (3.00, 2.29, 1.54) \cdot 10^{-3}\,\sigma^2/\tau$ for  the different volume fractions $(6, 9, 12)\,\%$ of Carbon Black. 

\subsection*{Alignment in shear flow}
\begin{figure}
    \centering
    \includegraphics[width=\linewidth]{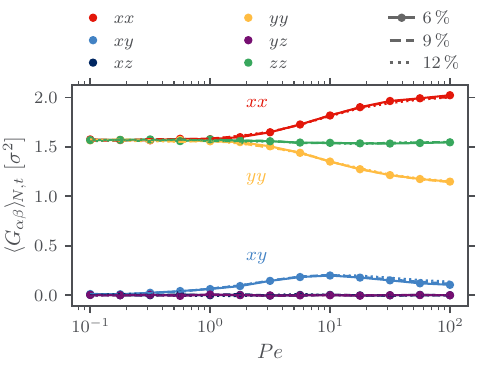}
    \caption{Time and ensemble averaged components of the gyration tensor $\left< G_{\alpha\beta} \right>_{N,t}$ plotted against the P\'eclet number $Pe$ for three volume fractions. As the shear rate grows, the $xx$ and $yy$-components increase, denoting an increased extent of the aggregates in flow direction and a decreased extent in gradient direction, respectively. The off-diagonal $xy$-component also indicates an emerging preferred orientation. Lines are guides to the eye.}
    \label{fig:gyration_tensor_components}
\end{figure}

Having established that Carbon Black aggregates can be significantly nonspherical, which hints to potentially interesting behavior in flow \cite{jeffery1922motion}, we set out to study their properties under shear in detail. As a first step, we assess the time and ensemble averaged components of the mean gyration tensor $\left< G_{\alpha\beta} \right>_{N,t}$ for varying strength of shear in \cref{fig:gyration_tensor_components}. For low shear rates, the off-diagonal entries vanish and the diagonal entries are equal. This marks the regime where diffusive transport dominates over advection from shear and where aggregate orientations are isotropic. For $Pe \gtrsim 1$, this behavior changes and we observe that the $xx$-component increases with increasing $Pe$, whereas the $yy$-component decreases. This signifies that the mean extent of aggregates becomes larger in the flow and smaller in the gradient direction. As the aggregates are stiff, this is not an effect of deformation, but solely of the orientation of the anisotropic aggregates. Interestingly, the measured values are almost identical for all studied volume fractions. Additionally, the $xy$-component increases, indicating a growing correlation between aggregates' primary particle positions along the flow and gradient directions and the emergence of a preferred orientation and alignment.

The alignment of aggregates can be analyzed explicitly using the nematic order parameter. For each of the aggregate axes, i.e., long, intermediate, and short, we define a tensor of orientational order via 
\begin{equation}\label{eq:qtensor}
    Q_{\alpha\beta}(t) = \frac{1}{2} \left< 3 e_{\alpha}(t) e_{\beta}(t) - \delta_{\alpha\beta} \right>_{N}\;,
\end{equation}
where $e_{\alpha}$ denotes a component of a normalized principal axis $\mathbf{e}$ of an aggregate and we average over all $N$ aggregates, $\delta_{\alpha\beta}$ is the Kronecker delta. The nematic tensor can be diagonalized, yielding its time-averaged largest eigenvalue $S$ as a measure for the alignment and the corresponding eigenvector $\mathbf{n}$ as the preferred direction of orientation, denoted director. The value of $S$ can express perfect alignment ($S=1$), the absence of alignment and preferential direction ($S=0$), or orthogonality with the director ($S=-0.5$). 

\begin{figure}
    \centering
    \includegraphics[width=\linewidth]{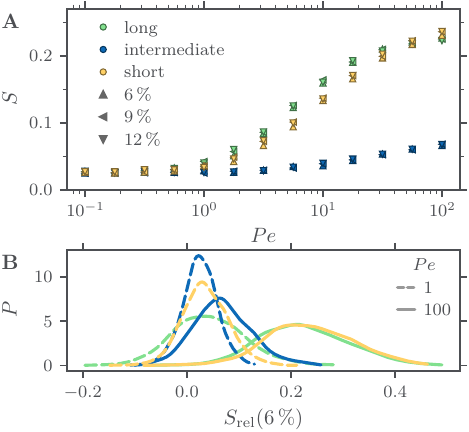}
    \caption{A: nematic order parameters $S$ of the three aggregate axes reveal increasing alignment of the long and short axes with increasing P\'eclet number $Pe$, as well as a less pronounced effect for the intermediate axis. 
    Uncertainties of $S$ from a block bootstrapping scheme would be smaller than the symbols. 
    B: probability distributions of $S_\mathrm{rel}$ at $\phi=6\,\%$ for $Pe = 0, 100$ show that the degree of alignment varies notably across aggregates. The kernel density estimates use a Gaussian kernel and Silverman's rule for the bandwidth determination.
    }
    \label{fig:NOP}
\end{figure}

We measure the time-averaged nematic order parameter $S$ for each aggregate axis and plot it against the P\'eclet number in \cref{fig:NOP}. At low $Pe$, finite-size effects lead to a small but nonzero apparent alignment. Beyond $Pe = 1$, the long and short axes show notable alignment effects. In the intermediary regime of $Pe \in [1, 32]$, the long axis is more aligned than the short axis. At higher shear, however, the short axis aligns to a similar degree. Interestingly, no effects of volume fraction can be observed, as the corresponding data points practically coincide. Apparently, the shear field rather than steric effects dominates the alignment. However, it should be noted that the values of the order parameter are lower than those typically found for nematic liquid crystals or rods in shear flow, where $S>0.5$ is common \cite{kuhnhold2016isotropic,dhont2006rod}. Therefore, the observed effect has to be considered a tendency of alignment. After all, Carbon Black aggregates are not simple cylindrical or ellipsoidal particles and exhibit a range of aspect ratios. On top, they perform tumbling motions in shear flow as shown further below, and are subject to collisions and diffusion.

To quantify the degree of alignment of individual aggregates, we calculate the relative alignment of each of their principal axes $\mathbf{e}$ with respect to the time-resolved director $\mathbf{n}(t)$ via
\begin{equation}
    S_\mathrm{rel} = \frac{1}{2} \left< 3 \left( \mathbf{e}(t)\cdot\mathbf{n}(t) \right)^2 - 1 \right>_t \;.
\end{equation}
The distributions of $S_\mathrm{rel}$ across aggregates are broad, which is depicted in \cref{fig:NOP}. The long axis has the widest distribution, which is matched by the short axis at high shear rates.
Overall, $S_\mathrm{rel}$ reflects the range of axis lengths and aspect ratios that our aggregate ensemble exhibits.

\begin{figure*}[t]
    \centering
    \includegraphics[width=\textwidth]{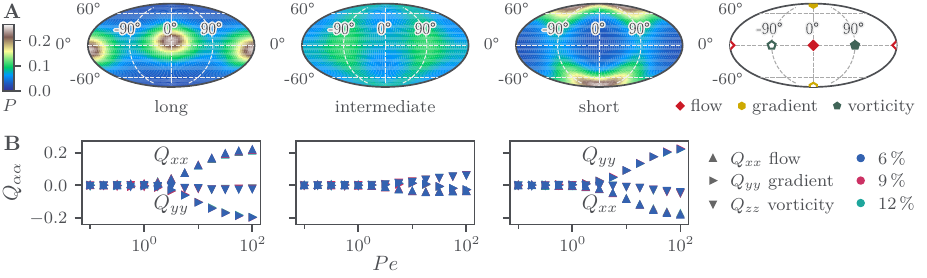}
    \caption{A: distribution of the long, intermediate, and short axes $\mathbf{e}$ of aggregates in shear flow with $Pe=100$, $\phi = 6\,\%$. The long and short axes show pronounced alignment close to the flow and vorticity directions, respectively. The fourth panel illustrates the shear flow axes within the Mollweide projection. Full symbols correspond to positive, empty symbols to negative directions. B: components of the $Q$-tensors of the three aggregate axes quantify the alignment with respect to the shear flow axes for varying P\'eclet number.}
    \label{fig:mollweide}
\end{figure*}

In \cref{fig:NOP}, the intermediate axis shows a much lower signal. Alignment effects seem to be weak. To study how this fits the overall picture, we visualize the orientations of the aggregates.  We unfold the three-dimensional unit sphere on which the axes reside by Mollweide projections, in analogy to globes and world maps. The flow direction is in the center of the projection at polar and azimuthal angles $(0^\circ, 0^\circ)$ and, due to head-tail-symmetry of the aggregate axes, also at $(0^\circ, \pm 180^\circ)$. The gradient direction is at the poles, the vorticity direction at $(0^\circ, \pm90^\circ)$. Bin populations are adjusted for the varying solid angle in the polar direction.

\Cref{fig:mollweide} shows the distributions of the three axes for all aggregates and times for $\phi=6\,\%$, $Pe=100$. For the long axis, we see a distinct maximum approximately in flow direction. This means that the long axis has a tendency to be parallel to the flow direction. Along the flow-vorticity plane, the probability is also increased. For the short axis, the characteristics are similar: it most often points in gradient direction or lies in the gradient-vorticity plane. The behavior of these two axes can be interpreted in the following way: an alignment effect from the shear field, as known from other elongated particles like rods \cite{dhont2006rod, jeffery1922motion}, influences the long axis. Thus, there is a mechanism that explicitly isolates and thus characterizes this axis. Regarding the short axis, particles are less prone to interactions and thus stabilized if their extent in the gradient direction is minimized. To this end, the short axis needs to be parallel to the gradient direction. 
For the intermediate axis, there is no mechanism that would distinguish it and single it out. Its maximum along the vorticity direction is the consequence of simultaneous alignment of the long and short axes with their respective directors.

Complementary, the diagonal entries of the $Q$-tensors introduced in \cref{eq:qtensor} can be interpreted. They encode the alignment of an aggregate axis with respect to the laboratory frame, i.e., the flow, gradient, and vorticity axis, and provide a more quantitative picture than the Mollweide plots. In \cref{fig:mollweide}, we present the components for varying P\'eclet number. Each aggregate axis is presented in a separate panel. For high shear rates, the long axis exhibits a positive $xx$-component, which implies a tendency to be parallel with the flow direction. Its negative $yy$-component denotes orthogonality with the gradient direction. This behavior, together with the analogous interpretation of the results for the intermediate and short axes, are consistent with the distributions of the axes shown as Mollweide plots. 

\begin{figure}
    \centering
    \includegraphics[]{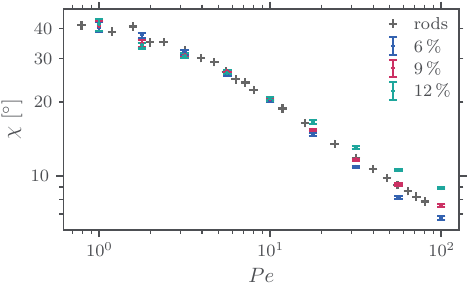}
    \caption{The orientation angle $\chi$ decreases with increasing P\'eclet number $Pe$. It is in agreement with data on isolated rods from Ref. \cite{winkler2004rod} that we extracted from the article. Error bars denote one-sigma intervals from bootstrapping on aggregates. 
    }
    \label{fig:orientation_angle_all}
\end{figure}

Upon close inspection of the population maxima of the long and short axes in the Mollweide projections in \cref{fig:mollweide}, it stands out that they are slightly shifted away from the shear flow axes. The long axis, for example, does not have its maximum at a polar angle of $\theta = 0^\circ$. Instead, it is found at an angle $\chi>0^\circ$ (for an azimuthal angle $\varphi\approx0^\circ$). We denote this characteristic angle between the long aggregate axis and the flow-vorticity plane the orientation angle \cite{aust1999structure,ripoll2006star} and calculate it via
\begin{equation}\label{eq:orientation}
    \tan(2\chi) =  \frac{2\,\langle G_{xy}\rangle_{N,t}}{\langle G_{xx}\rangle_{N,t} - \langle G_{yy}\rangle_{N,t}}\;.
\end{equation}
For the lowest shear rates, \cref{fig:orientation_angle_all} shows that $\chi$ is around $40^\circ$, approaching the $45^\circ$ expected for rods \cite{dhont2006rod,winkler2004rod}. For even lower shear rates with $Pe<1$, the computation becomes numerically unstable, as both numerator and denominator in \cref{eq:orientation} vanish (see \cref{fig:gyration_tensor_components}). Together with the weak alignment measured by the nematic order parameter (\cref{fig:NOP}), this means that the concept of the orientation angle is no longer meaningful. With increasing P\'eclet number, the orientation angle decreases. At the highest observed shear rates, small effects of volume fraction can be recognized, where higher volume fractions mean higher orientation angles. We compare our results to simulations of isolated rods with aspect ratio $15$  in a hydrodynamic solvent \cite{winkler2004rod, dhont2006rod}. The orientation angles $\chi$ of these rods are included in \cref{fig:orientation_angle_all}, assuming that the reduced shear rate coincides with our P\'eclet number. Our data on aggregates is in excellent agreement with the simulations of rods over a wide range of shear rates.

\subsection*{Tumbling}
Elongated ellipsoids and bodies of revolution rotate continuously in dilute, sheared suspensions, following Jeffery orbits \cite{jeffery1922motion,bretherton1962motion}. Because certain orientations are dynamically more stable, the previously reported alignment and the orientation angle occur. This tumbling motion can also be expected for Carbon Black aggregates due to their anisotropy, even though the behavior is in general more complex in the absence of axial symmetry \cite{hinch1979rotation,xu2014shear,trofa2020rheology}. 

To measure the relevant time scale of this tumbling motion, we start out with the time traces $\Delta G_{\alpha\alpha}(t) = G_{\alpha\alpha}(t) - \langle G_{\alpha\alpha}\rangle_{t}$, which are related to the extent of a single aggregate in the flow ($xx$) or gradient ($yy$) direction. A time shift $\Delta t$ for which $\Delta G_{xx}$ and $\Delta G_{yy}$ become most similar to each other corresponds to the duration of a $90^\circ$ rotation of a tumbling aggregate. To extract this time shift, we calculate the cross-correlation function of the flow and gradient components of the gyration tensor
\begin{equation}\label{eq:tumble_ccf}
    C^\ast_{xy}(\Delta t) = \frac{\langle \Delta G_{yy}(t) \Delta G_{xx}(t + \Delta t) \rangle_{t} }{\sqrt{\langle \Delta G^2_{yy}(t) \rangle_{t}\langle \Delta G^2_{xx}(t) \rangle_{t}}}\;.
\end{equation}
for each aggregate individually and for each P\'eclet number. To define one effective tumbling time scale per $Pe$, $C^\ast_{xy}$ is averaged over all $N$ aggregates, i.e., $C_{xy}(\Delta t) = \langle C^\ast_{xy}(\Delta t)\rangle_N$. 

\begin{figure}
    \centering
    \includegraphics[]{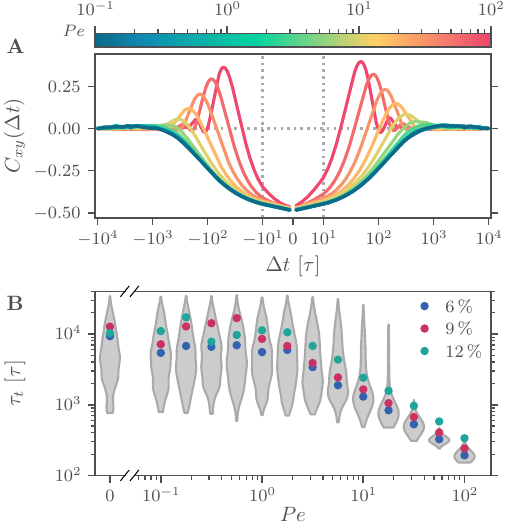}
    \caption{A: the cross-correlation function of the flow and gradient components of the gyration tensor $C_{xy}(\Delta t)$ captures the tumbling behavior for varying $Pe$. The first maxima at positive lag times relate to the characteristic tumbling time.
    The lag time is shown on a symmetric logarithmic scale, linear for $\Delta t/\tau \in [-10, 10]$.
    B: The tumbling time $\tau_t$ decreases for increasing P\'eclet number $Pe$ beyond $Pe>1$. At about $Pe<1$, diffusion dominates and no tumbling can be seen. To resolve the tumbling of single aggregates, violin plots of the distribution of tumbling times of individual aggregates are added for $\phi = 6\,\%$.
    }
    \label{fig:tumbling_ccf}
\end{figure}

For several high P\'eclet numbers, the cross-correlation functions $C_{xy}(\Delta t)$ shown in \cref{fig:tumbling_ccf} exhibit prominent maxima at positive lag times $\Delta t_\mathrm{max}$. As these correspond to the aforementioned $90^\circ$ rotation, we define the tumbling time as $4\Delta t_\mathrm{max} = \tau_t$. Higher-order maxima exist for the strongest observed shear. From purely monofrequent tumbling, one would expect a periodic correlation function. However, due to the range of aspect ratios of the aggregates, thermal noise, and interactions, the aggregate-wise correlation functions $C^\ast_{xy}(\Delta t)$ dephase and decrease in magnitude. It stands out that the correlation functions are almost symmetrical in lag time, which is not the case for flexible polymers. Asymmetry appears to be a signature of polymer flexibility and tank-treading \cite{chen2013tumbling,sablic2017deciphering,huang2011tumbling}. For star polymers, reducing the number of arms deforms the curve towards lag time symmetry \cite{peng2024unveiling}. 

\Cref{fig:tumbling_ccf} also presents the characteristic tumbling time $\tau_t$ plotted against the P\'eclet number. The tumbling frequency is approximately constant until $Pe\approx 1$ as the motion is dominated by diffusion. 
Beyond, $\tau_t$ decreases strongly and appears to follow a power law. 
While the shown two decades do not warrant a reliable fit of an exponent, other authors have observed similar behavior for semiflexible rods \cite{nikoubashman2017equilibrium, winkler2006semiflexible}, helices \cite{li2021tumbling}, polymers and polymer rings \cite{schroeder2005characteristic,lang2014dynamics,saha2012tumbling,xu2014shear,winkler2014dynamical} with various exponents, and Jeffery orbits \cite{jeffery1922motion} expect $\tau_T \propto \dot\gamma^{-1}$.

For the highest P\'eclet numbers, higher volume fractions are consistently associated with slower tumbling. We measure that for $Pe\geq1$, the tumbling periods at $\phi = 9\,\%~(12\,\%)$ are on average $23\,\%~(90\,\%)$ faster than at $\phi = 6\,\%$, which we attribute to excluded volume interactions. Overall, the tumbling time appears to be especially sensitive to density effects.

As $\tau_t$ is derived from the ensemble-averaged cross-correlation $C_{xy}(t)$, it reflects a characteristic time scale of the entire system depending on the shear rate. To resolve the motion of individual aggregates, we derive tumbling times $\tau^\ast_t$ from their correlation functions $C^\ast_{xy}(t)$ in an analogous manner. This comes at the cost of a worse signal-to-noise ratio. The distributions are included in \cref{fig:tumbling_ccf} as violin plots, whose widths reflect kernel density estimates of the tumbling-time distributions. At low shear rates, $\tau^\ast_t$ exhibits a large variance. In this regime, there is no strong alignment effect and since we rely on the cross-correlation function of the flow and gradient components of the gyration tensor, this renders the approach less effective for both $\tau_t$ and $\tau^\ast_t$. At higher shear rates, the distributions of $\tau^\ast_t$ become narrower, which signifies that most aggregates tumble with similar frequencies.

Rotation dynamics of aggregates are therefore initially dominated by diffusion and do not depend on the shear rate. Then, the minimal P\'eclet number for which shear leads to a change in behavior is similar for tumbling and for alignment. The tumbling speeds up with increasing shear strength, likely following a power law.

\subsection*{Viscosity}
\begin{figure}[t]
    \centering
    \includegraphics[width=\linewidth]{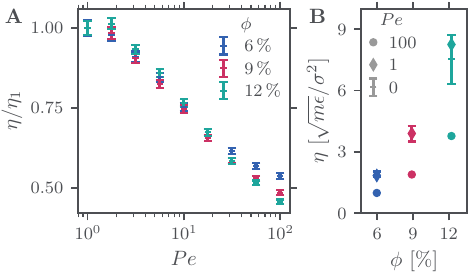}
    \caption{A: the normalized viscosity $\eta/\eta_1$ decreases with increasing P\'eclet number $Pe$. B: higher volume fractions $\phi$ lead to higher viscosity. We combine results from the Green-Kubo relation for $Pe=0$ and from simulations under shear for $Pe=1, 100$. Error bars denote the standard error, but are smaller than the symbols for $Pe=1, 100$ and are thus omitted. }
    \label{fig:viscosity}
\end{figure}
Next, we will analyze whether the observed alignment effects reveal themselves in the viscosity. To this end, we measure the stress via
\begin{equation}
    \sigma_{xy} = -\frac{1}{V} \sum_i^{N \cdot N_p} \left( r_{i,x} F_{i,y} + m v_{i,x} v_{i,y}\right)\;,
\end{equation}
where $r_{ix}$ denotes the $x$-component of the position of primary particle $i$, $F_{iy}$ is the $y$-component of the force acting on it, and $v_{ix}$ is the $x$-component of the velocity. The sum runs over all primary particles. Then, the viscosity $\eta$ acts as the proportionality constant relating the time-averaged stress to the applied shear rate,
\begin{equation}\label{eq:nemd}
    \eta = \left<\sigma_{xy}\right>_t/\dot\gamma\;,
\end{equation}
under steady shear.
Alternatively, in the quiescent state, we can make use of the Green-Kubo formalism \cite{zwanzig1965time} via
\begin{equation}\label{eq:green-kubo}
    \eta = \frac{V}{k_\mathrm{B}T} \int\limits_0^\infty \mathrm{d}s \langle \sigma_{xy}(t) \sigma_{xy}(t+s) \rangle_{t}\;.
\end{equation}
The viscosity results are shown in \cref{fig:viscosity}. If plotted against the volume fraction, it becomes evident that higher densities lead to higher viscosities. The Green-Kubo results with $Pe=0$ agree well with the viscosity measured under shear flow for $Pe=1$. We know the latter with much higher precision (maximal relative uncertainty of $2\,\%$ for $Pe=1$ versus $16\,\%$ for $Pe=0$). Therefore, we normalize the viscosity under shear flow with $\eta(Pe=1) = \eta_1$ to capture relative changes.

With increasing P\'eclet number, this normalized viscosity decreases notably, reducing by approximately $50\,\%$ for the highest shear rates. Thus, there is a clear shear thinning effect. We explain this behavior with the alignment of the Carbon Black aggregates, concerning both the long and short axes. This argument is in agreement with prior studies on other elongated particles or polymers \cite{nikoubashman2017equilibrium,xu2014shear}. For most $Pe$, the normalization leads to coinciding curves for the different volume fractions. Only for the highest P\'eclet numbers, the shear thinning becomes stronger for higher volume fractions. Thus, density effects do not only show up in the absolute viscosity, but also in the strength of shear thinning.

In general, the viscosity of Carbon Black can be more complex if attractions \cite{vermant2005flow,richards2023review} are considered, which lead to clusters of aggregates, the agglomerates. Additionally, hydrodynamic interactions could be included, which have been observed to possibly change the rheological response of spherical colloids to shear \cite{brady2001computer,varga2019hydrodynamics}.

\section{Conclusion}
We have investigated the non-equilibrium steady states of suspensions of fractal Carbon Black aggregates in simple shear flow. Addressing the context of conductive suspensions, we focused on volume fractions around the connectivity percolation threshold. Although the aggregates are assembled in an intrinsically isotropic aggregation process, which we emulated with diffusion-limited aggregation, their structure is markedly anisotropic. 
Accordingly, the aggregates behave similarly to ellipsoidal or rod-like particles under shear flow. At P\'eclet numbers of about $Pe = 1$, the aggregates' longest and shortest axes start to align with the flow and gradient directions, respectively. They undergo a tumbling motion that speeds up with applied shear rate, and the suspensions exhibit shear-thinning rheology. Effects of density are observed for the viscosity, the orientation angle, and for the tumbling times, whereas the degree of alignment is unaffected.

Future extensive simulations are expected to elucidate how shear flow influences transient particle networks that support electrical conductivity. Especially, it remains unclear whether anisotropic cluster formation under shear or aggregate-level alignment dominates conductivity characteristics.

The findings presented here have direct relevance for electrically conductive suspensions, nanocomposite precursors, and other applications in which Carbon Black aggregates are used. When shear flows are involved, the anisotropic nature of the aggregates likely cannot be neglected. Instead, it can be exploited to induce and tune directionally dependent material properties and functional behavior.

\begin{acknowledgments}
The authors thank Anja Seegebrecht for constructive criticism of the manuscript. The authors acknowledge support by the state of Baden-Württemberg through bwHPC and the German Research Foundation (DFG) through grant no INST 39/963-1 FUGG (bwForCluster NEMO). 
The authors would like to thank the state of Baden-Württemberg for its support of bwHPC and the German Research Foundation (DFG) for funding under "Project number 455622343" (bwForCluster NEMO 2).
Funded by the Deutsche Forschungsgemeinschaft (DFG, German Research Foundation) - 531007218. Funded by the Deutsche Forschungsgemeinschaft (DFG, German Research Foundation) under Germany's Excellence Strategy – EXC-2193/1 – 390951807. The use of \texttt{prettypyplot} \cite{prettypyplot} for data visualization is gratefully acknowledged.
\end{acknowledgments}

\section*{Conflict of Interest}
The authors have no conflicts to disclose.

\section*{Data availability}
The data that support the findings of this study are available
from the corresponding author upon reasonable request.


\begin{thebibliography}{62}%
	\makeatletter
	\providecommand \@ifxundefined [1]{%
		\@ifx{#1\undefined}
	}%
	\providecommand \@ifnum [1]{%
		\ifnum #1\expandafter \@firstoftwo
		\else \expandafter \@secondoftwo
		\fi
	}%
	\providecommand \@ifx [1]{%
		\ifx #1\expandafter \@firstoftwo
		\else \expandafter \@secondoftwo
		\fi
	}%
	\providecommand \natexlab [1]{#1}%
	\providecommand \enquote  [1]{``#1''}%
	\providecommand \bibnamefont  [1]{#1}%
	\providecommand \bibfnamefont [1]{#1}%
	\providecommand \citenamefont [1]{#1}%
	\providecommand \href@noop [0]{\@secondoftwo}%
	\providecommand \href [0]{\begingroup \@sanitize@url \@href}%
	\providecommand \@href[1]{\@@startlink{#1}\@@href}%
	\providecommand \@@href[1]{\endgroup#1\@@endlink}%
	\providecommand \@sanitize@url [0]{\catcode `\\12\catcode `\$12\catcode
		`\&12\catcode `\#12\catcode `\^12\catcode `\_12\catcode `\%12\relax}%
	\providecommand \@@startlink[1]{}%
	\providecommand \@@endlink[0]{}%
	\providecommand \url  [0]{\begingroup\@sanitize@url \@url }%
	\providecommand \@url [1]{\endgroup\@href {#1}{\urlprefix }}%
	\providecommand \urlprefix  [0]{URL }%
	\providecommand \Eprint [0]{\href }%
	\providecommand \doibase [0]{https://doi.org/}%
	\providecommand \selectlanguage [0]{\@gobble}%
	\providecommand \bibinfo  [0]{\@secondoftwo}%
	\providecommand \bibfield  [0]{\@secondoftwo}%
	\providecommand \translation [1]{[#1]}%
	\providecommand \BibitemOpen [0]{}%
	\providecommand \bibitemStop [0]{}%
	\providecommand \bibitemNoStop [0]{.\EOS\space}%
	\providecommand \EOS [0]{\spacefactor3000\relax}%
	\providecommand \BibitemShut  [1]{\csname bibitem#1\endcsname}%
	\let\auto@bib@innerbib\@empty
	%</preamble>
	\bibitem [{\citenamefont {Loffredo}\ \emph {et~al.}(2009)\citenamefont
		{Loffredo}, \citenamefont {Del~Mauro}, \citenamefont {Burrasca},
		\citenamefont {La~Ferrara}, \citenamefont {Quercia}, \citenamefont {Massera},
		\citenamefont {Di~Francia},\ and\ \citenamefont
		{Della~Sala}}]{loffredo2009ink}%
	\BibitemOpen
	\bibfield  {author} {\bibinfo {author} {\bibfnamefont {F.}~\bibnamefont
			{Loffredo}}, \bibinfo {author} {\bibfnamefont {A.~D.~G.}\ \bibnamefont
			{Del~Mauro}}, \bibinfo {author} {\bibfnamefont {G.}~\bibnamefont {Burrasca}},
		\bibinfo {author} {\bibfnamefont {V.}~\bibnamefont {La~Ferrara}}, \bibinfo
		{author} {\bibfnamefont {L.}~\bibnamefont {Quercia}}, \bibinfo {author}
		{\bibfnamefont {E.}~\bibnamefont {Massera}}, \bibinfo {author} {\bibfnamefont
			{G.}~\bibnamefont {Di~Francia}},\ and\ \bibinfo {author} {\bibfnamefont
			{D.}~\bibnamefont {Della~Sala}},\ }\href
	{https://doi.org/10.1016/j.snb.2009.09.024} {\bibfield  {journal} {\bibinfo
			{journal} {Sens. Actuators, B}\ }\textbf {\bibinfo {volume} {143}},\ \bibinfo
		{pages} {421} (\bibinfo {year} {2009})}\BibitemShut {NoStop}%
	\bibitem [{\citenamefont {Cui}(2016)}]{cui2016printed}%
	\BibitemOpen
	\bibfield  {author} {\bibinfo {author} {\bibfnamefont {Z.}~\bibnamefont
			{Cui}},\ }\href {https://doi.org/10.1002/9781118920954} {\emph {\bibinfo
			{title} {Printed electronics: materials, technologies and applications}}}\
	(\bibinfo  {publisher} {John Wiley \& Sons},\ \bibinfo {year}
	{2016})\BibitemShut {NoStop}%
	\bibitem [{\citenamefont {Camargo}\ \emph {et~al.}(2021)\citenamefont
		{Camargo}, \citenamefont {Orzari}, \citenamefont {Araujo}, \citenamefont
		{de~Oliveira}, \citenamefont {Kalinke}, \citenamefont {Rocha}, \citenamefont
		{dos Santos}, \citenamefont {Takeuchi}, \citenamefont {Munoz}, \citenamefont
		{Bonacin} \emph {et~al.}}]{camargo2021development}%
	\BibitemOpen
	\bibfield  {author} {\bibinfo {author} {\bibfnamefont {J.~R.}\ \bibnamefont
			{Camargo}}, \bibinfo {author} {\bibfnamefont {L.~O.}\ \bibnamefont {Orzari}},
		\bibinfo {author} {\bibfnamefont {D.~A.~G.}\ \bibnamefont {Araujo}}, \bibinfo
		{author} {\bibfnamefont {P.~R.}\ \bibnamefont {de~Oliveira}}, \bibinfo
		{author} {\bibfnamefont {C.}~\bibnamefont {Kalinke}}, \bibinfo {author}
		{\bibfnamefont {D.~P.}\ \bibnamefont {Rocha}}, \bibinfo {author}
		{\bibfnamefont {A.~L.}\ \bibnamefont {dos Santos}}, \bibinfo {author}
		{\bibfnamefont {R.~M.}\ \bibnamefont {Takeuchi}}, \bibinfo {author}
		{\bibfnamefont {R.~A.~A.}\ \bibnamefont {Munoz}}, \bibinfo {author}
		{\bibfnamefont {J.~A.}\ \bibnamefont {Bonacin}}, \emph {et~al.},\ }\href
	{https://doi.org/10.1016/j.microc.2021.105998} {\bibfield  {journal}
		{\bibinfo  {journal} {Microchem. J.}\ }\textbf {\bibinfo {volume} {164}},\
		\bibinfo {pages} {105998} (\bibinfo {year} {2021})}\BibitemShut {NoStop}%
	\bibitem [{\citenamefont {Cummins}\ and\ \citenamefont
		{Desmulliez}(2012)}]{cummins2012inkjet}%
	\BibitemOpen
	\bibfield  {author} {\bibinfo {author} {\bibfnamefont {G.}~\bibnamefont
			{Cummins}}\ and\ \bibinfo {author} {\bibfnamefont {M.~P.}\ \bibnamefont
			{Desmulliez}},\ }\href {https://doi.org/10.1108/03056121211280413} {\bibfield
		{journal} {\bibinfo  {journal} {Circuit World}\ }\textbf {\bibinfo {volume}
			{38}},\ \bibinfo {pages} {193} (\bibinfo {year} {2012})}\BibitemShut
	{NoStop}%
	\bibitem [{\citenamefont {Schmidt}\ \emph {et~al.}(2024)\citenamefont
		{Schmidt}, \citenamefont {Kraus},\ and\ \citenamefont
		{Gonz{\'a}lez-Garc{\'\i}a}}]{schmidt2024electrofluids}%
	\BibitemOpen
	\bibfield  {author} {\bibinfo {author} {\bibfnamefont {D.~S.}\ \bibnamefont
			{Schmidt}}, \bibinfo {author} {\bibfnamefont {T.}~\bibnamefont {Kraus}},\
		and\ \bibinfo {author} {\bibfnamefont {L.}~\bibnamefont
			{Gonz{\'a}lez-Garc{\'\i}a}},\ }\href {https://doi.org/10.1021/acsami.4c07230}
	{\bibfield  {journal} {\bibinfo  {journal} {ACS Appl. Mater. Interfaces}\
		}\textbf {\bibinfo {volume} {16}},\ \bibinfo {pages} {43942} (\bibinfo {year}
		{2024})}\BibitemShut {NoStop}%
	\bibitem [{\citenamefont {Presser}\ \emph {et~al.}(2012)\citenamefont
		{Presser}, \citenamefont {Dennison}, \citenamefont {Campos}, \citenamefont
		{Knehr}, \citenamefont {Kumbur},\ and\ \citenamefont
		{Gogotsi}}]{presser2012electrochemical}%
	\BibitemOpen
	\bibfield  {author} {\bibinfo {author} {\bibfnamefont {V.}~\bibnamefont
			{Presser}}, \bibinfo {author} {\bibfnamefont {C.~R.}\ \bibnamefont
			{Dennison}}, \bibinfo {author} {\bibfnamefont {J.}~\bibnamefont {Campos}},
		\bibinfo {author} {\bibfnamefont {K.~W.}\ \bibnamefont {Knehr}}, \bibinfo
		{author} {\bibfnamefont {E.~C.}\ \bibnamefont {Kumbur}},\ and\ \bibinfo
		{author} {\bibfnamefont {Y.}~\bibnamefont {Gogotsi}},\ }\href
	{https://doi.org/10.1002/aenm.201100768} {\bibfield  {journal} {\bibinfo
			{journal} {Adv. Energy Mater.}\ }\textbf {\bibinfo {volume} {2}},\ \bibinfo
		{pages} {895} (\bibinfo {year} {2012})}\BibitemShut {NoStop}%
	\bibitem [{\citenamefont {Campos}\ \emph {et~al.}(2013)\citenamefont {Campos},
		\citenamefont {Beidaghi}, \citenamefont {Hatzell}, \citenamefont {Dennison},
		\citenamefont {Musci}, \citenamefont {Presser}, \citenamefont {Kumbur},\ and\
		\citenamefont {Gogotsi}}]{campos2013investigation}%
	\BibitemOpen
	\bibfield  {author} {\bibinfo {author} {\bibfnamefont {J.~W.}\ \bibnamefont
			{Campos}}, \bibinfo {author} {\bibfnamefont {M.}~\bibnamefont {Beidaghi}},
		\bibinfo {author} {\bibfnamefont {K.~B.}\ \bibnamefont {Hatzell}}, \bibinfo
		{author} {\bibfnamefont {C.~R.}\ \bibnamefont {Dennison}}, \bibinfo {author}
		{\bibfnamefont {B.}~\bibnamefont {Musci}}, \bibinfo {author} {\bibfnamefont
			{V.}~\bibnamefont {Presser}}, \bibinfo {author} {\bibfnamefont {E.~C.}\
			\bibnamefont {Kumbur}},\ and\ \bibinfo {author} {\bibfnamefont
			{Y.}~\bibnamefont {Gogotsi}},\ }\href
	{https://doi.org/10.1016/j.electacta.2013.03.037} {\bibfield  {journal}
		{\bibinfo  {journal} {Electrochim. Acta}\ }\textbf {\bibinfo {volume} {98}},\
		\bibinfo {pages} {123} (\bibinfo {year} {2013})}\BibitemShut {NoStop}%
	\bibitem [{\citenamefont {Zhang}\ \emph {et~al.}(2021)\citenamefont {Zhang},
		\citenamefont {Ma}, \citenamefont {Wu}, \citenamefont {Sun}, \citenamefont
		{Wang}, \citenamefont {Li},\ and\ \citenamefont {Waite}}]{zhang2021flow}%
	\BibitemOpen
	\bibfield  {author} {\bibinfo {author} {\bibfnamefont {C.}~\bibnamefont
			{Zhang}}, \bibinfo {author} {\bibfnamefont {J.}~\bibnamefont {Ma}}, \bibinfo
		{author} {\bibfnamefont {L.}~\bibnamefont {Wu}}, \bibinfo {author}
		{\bibfnamefont {J.}~\bibnamefont {Sun}}, \bibinfo {author} {\bibfnamefont
			{L.}~\bibnamefont {Wang}}, \bibinfo {author} {\bibfnamefont {T.}~\bibnamefont
			{Li}},\ and\ \bibinfo {author} {\bibfnamefont {T.~D.}\ \bibnamefont
			{Waite}},\ }\href {https://doi.org/10.1021/acs.est.0c06552} {\bibfield
		{journal} {\bibinfo  {journal} {Environ. Sci. Technol.}\ }\textbf {\bibinfo
			{volume} {55}},\ \bibinfo {pages} {4243} (\bibinfo {year}
		{2021})}\BibitemShut {NoStop}%
	\bibitem [{\citenamefont {Alfonso}\ \emph {et~al.}(2021)\citenamefont
		{Alfonso}, \citenamefont {Parant}, \citenamefont {Yuan}, \citenamefont
		{Neri}, \citenamefont {Laurichesse}, \citenamefont {Kampioti}, \citenamefont
		{Colin},\ and\ \citenamefont {Poulin}}]{alfonso2021highly}%
	\BibitemOpen
	\bibfield  {author} {\bibinfo {author} {\bibfnamefont {M.~S.}\ \bibnamefont
			{Alfonso}}, \bibinfo {author} {\bibfnamefont {H.}~\bibnamefont {Parant}},
		\bibinfo {author} {\bibfnamefont {J.}~\bibnamefont {Yuan}}, \bibinfo {author}
		{\bibfnamefont {W.}~\bibnamefont {Neri}}, \bibinfo {author} {\bibfnamefont
			{E.}~\bibnamefont {Laurichesse}}, \bibinfo {author} {\bibfnamefont
			{K.}~\bibnamefont {Kampioti}}, \bibinfo {author} {\bibfnamefont
			{A.}~\bibnamefont {Colin}},\ and\ \bibinfo {author} {\bibfnamefont
			{P.}~\bibnamefont {Poulin}},\ }\bibfield  {journal} {\bibinfo  {journal}
		{Iscience}\ }\textbf {\bibinfo {volume} {24}},\ \href
	{https://doi.org/10.1016/j.isci.2021.102456} {10.1016/j.isci.2021.102456}
	(\bibinfo {year} {2021})\BibitemShut {NoStop}%
	\bibitem [{\citenamefont {Wonisch}\ \emph {et~al.}(2011)\citenamefont
		{Wonisch}, \citenamefont {Polfer}, \citenamefont {Kraft}, \citenamefont
		{Dellert}, \citenamefont {Heunisch},\ and\ \citenamefont
		{Roosen}}]{wonisch2011comprehensive}%
	\BibitemOpen
	\bibfield  {author} {\bibinfo {author} {\bibfnamefont {A.}~\bibnamefont
			{Wonisch}}, \bibinfo {author} {\bibfnamefont {P.}~\bibnamefont {Polfer}},
		\bibinfo {author} {\bibfnamefont {T.}~\bibnamefont {Kraft}}, \bibinfo
		{author} {\bibfnamefont {A.}~\bibnamefont {Dellert}}, \bibinfo {author}
		{\bibfnamefont {A.}~\bibnamefont {Heunisch}},\ and\ \bibinfo {author}
		{\bibfnamefont {A.}~\bibnamefont {Roosen}},\ }\href
	{https://doi.org/10.1111/j.1551-2916.2010.04358.x} {\bibfield  {journal}
		{\bibinfo  {journal} {J. Am. Ceram. Soc.}\ }\textbf {\bibinfo {volume}
			{94}},\ \bibinfo {pages} {2053} (\bibinfo {year} {2011})}\BibitemShut
	{NoStop}%
	\bibitem [{\citenamefont {Choi}\ \emph {et~al.}(2019)\citenamefont {Choi},
		\citenamefont {Han}, \citenamefont {Kim}, \citenamefont {Hyeon},\ and\
		\citenamefont {Kim}}]{choi2019high}%
	\BibitemOpen
	\bibfield  {author} {\bibinfo {author} {\bibfnamefont {S.}~\bibnamefont
			{Choi}}, \bibinfo {author} {\bibfnamefont {S.~I.}\ \bibnamefont {Han}},
		\bibinfo {author} {\bibfnamefont {D.}~\bibnamefont {Kim}}, \bibinfo {author}
		{\bibfnamefont {T.}~\bibnamefont {Hyeon}},\ and\ \bibinfo {author}
		{\bibfnamefont {D.-H.}\ \bibnamefont {Kim}},\ }\href
	{https://doi.org/10.1039/C8CS00706C} {\bibfield  {journal} {\bibinfo
			{journal} {Chem. Soc. Rev.}\ }\textbf {\bibinfo {volume} {48}},\ \bibinfo
		{pages} {1566} (\bibinfo {year} {2019})}\BibitemShut {NoStop}%
	\bibitem [{\citenamefont {Li}\ \emph {et~al.}(2019)\citenamefont {Li},
		\citenamefont {Huang}, \citenamefont {Zeng}, \citenamefont {Li},
		\citenamefont {Tian}, \citenamefont {Fu}, \citenamefont {Wang},\ and\
		\citenamefont {Zhong}}]{li2019review}%
	\BibitemOpen
	\bibfield  {author} {\bibinfo {author} {\bibfnamefont {Y.}~\bibnamefont
			{Li}}, \bibinfo {author} {\bibfnamefont {X.}~\bibnamefont {Huang}}, \bibinfo
		{author} {\bibfnamefont {L.}~\bibnamefont {Zeng}}, \bibinfo {author}
		{\bibfnamefont {R.}~\bibnamefont {Li}}, \bibinfo {author} {\bibfnamefont
			{H.}~\bibnamefont {Tian}}, \bibinfo {author} {\bibfnamefont {X.}~\bibnamefont
			{Fu}}, \bibinfo {author} {\bibfnamefont {Y.}~\bibnamefont {Wang}},\ and\
		\bibinfo {author} {\bibfnamefont {W.-H.}\ \bibnamefont {Zhong}},\ }\href
	{https://doi.org/10.1007/s10853-018-3006-9} {\bibfield  {journal} {\bibinfo
			{journal} {J. Mater. Sci.}\ }\textbf {\bibinfo {volume} {54}},\ \bibinfo
		{pages} {1036} (\bibinfo {year} {2019})}\BibitemShut {NoStop}%
	\bibitem [{\citenamefont {Kanoun}\ \emph {et~al.}(2021)\citenamefont {Kanoun},
		\citenamefont {Bouhamed}, \citenamefont {Ramalingame}, \citenamefont
		{Bautista-Quijano}, \citenamefont {Rajendran},\ and\ \citenamefont
		{Al-Hamry}}]{kanoun2021review}%
	\BibitemOpen
	\bibfield  {author} {\bibinfo {author} {\bibfnamefont {O.}~\bibnamefont
			{Kanoun}}, \bibinfo {author} {\bibfnamefont {A.}~\bibnamefont {Bouhamed}},
		\bibinfo {author} {\bibfnamefont {R.}~\bibnamefont {Ramalingame}}, \bibinfo
		{author} {\bibfnamefont {J.~R.}\ \bibnamefont {Bautista-Quijano}}, \bibinfo
		{author} {\bibfnamefont {D.}~\bibnamefont {Rajendran}},\ and\ \bibinfo
		{author} {\bibfnamefont {A.}~\bibnamefont {Al-Hamry}},\ }\href
	{https://doi.org/10.3390/s21020341} {\bibfield  {journal} {\bibinfo
			{journal} {Sensors}\ }\textbf {\bibinfo {volume} {21}},\ \bibinfo {pages}
		{341} (\bibinfo {year} {2021})}\BibitemShut {NoStop}%
	\bibitem [{\citenamefont {Thomassin}\ \emph {et~al.}(2013)\citenamefont
		{Thomassin}, \citenamefont {Jerome}, \citenamefont {Pardoen}, \citenamefont
		{Bailly}, \citenamefont {Huynen},\ and\ \citenamefont
		{Detrembleur}}]{thomassin2013polymer}%
	\BibitemOpen
	\bibfield  {author} {\bibinfo {author} {\bibfnamefont {J.-M.}\ \bibnamefont
			{Thomassin}}, \bibinfo {author} {\bibfnamefont {C.}~\bibnamefont {Jerome}},
		\bibinfo {author} {\bibfnamefont {T.}~\bibnamefont {Pardoen}}, \bibinfo
		{author} {\bibfnamefont {C.}~\bibnamefont {Bailly}}, \bibinfo {author}
		{\bibfnamefont {I.}~\bibnamefont {Huynen}},\ and\ \bibinfo {author}
		{\bibfnamefont {C.}~\bibnamefont {Detrembleur}},\ }\href
	{https://doi.org/10.1016/j.mser.2013.06.001} {\bibfield  {journal} {\bibinfo
			{journal} {Mater. Sci. Eng. R Rep.}\ }\textbf {\bibinfo {volume} {74}},\
		\bibinfo {pages} {211} (\bibinfo {year} {2013})}\BibitemShut {NoStop}%
	\bibitem [{\citenamefont {Tan}\ \emph {et~al.}(2019)\citenamefont {Tan},
		\citenamefont {An}, \citenamefont {Chua},\ and\ \citenamefont
		{Tran}}]{tan2019metallic}%
	\BibitemOpen
	\bibfield  {author} {\bibinfo {author} {\bibfnamefont {H.~W.}\ \bibnamefont
			{Tan}}, \bibinfo {author} {\bibfnamefont {J.}~\bibnamefont {An}}, \bibinfo
		{author} {\bibfnamefont {C.~K.}\ \bibnamefont {Chua}},\ and\ \bibinfo
		{author} {\bibfnamefont {T.}~\bibnamefont {Tran}},\ }\href
	{https://doi.org/10.1002/aelm.201800831} {\bibfield  {journal} {\bibinfo
			{journal} {Adv. Electron. Mater.}\ }\textbf {\bibinfo {volume} {5}},\
		\bibinfo {pages} {1800831} (\bibinfo {year} {2019})}\BibitemShut {NoStop}%
	\bibitem [{\citenamefont {Watson}\ and\ \citenamefont
		{Valberg}(2001)}]{watson2001carbon}%
	\BibitemOpen
	\bibfield  {author} {\bibinfo {author} {\bibfnamefont {A.~Y.}\ \bibnamefont
			{Watson}}\ and\ \bibinfo {author} {\bibfnamefont {P.~A.}\ \bibnamefont
			{Valberg}},\ }\href {https://doi.org/10.1080/15298660108984625} {\bibfield
		{journal} {\bibinfo  {journal} {AIHAJ-American Industrial Hygiene
				Association}\ }\textbf {\bibinfo {volume} {62}},\ \bibinfo {pages} {218}
		(\bibinfo {year} {2001})}\BibitemShut {NoStop}%
	\bibitem [{\citenamefont {Kl{\"u}ppel}(2014)}]{kluppel2014role}%
	\BibitemOpen
	\bibfield  {author} {\bibinfo {author} {\bibfnamefont {M.}~\bibnamefont
			{Kl{\"u}ppel}},\ }in\ \href@noop {} {\emph {\bibinfo {booktitle}
			{Filler-reinforced elastomers scanning force microscopy}}}\ (\bibinfo
	{publisher} {Springer},\ \bibinfo {year} {2014})\ pp.\ \bibinfo {pages}
	{1--86}\BibitemShut {NoStop}%
	\bibitem [{\citenamefont {Coupette}\ \emph {et~al.}(2021)\citenamefont
		{Coupette}, \citenamefont {Zhang}, \citenamefont {Kuttich}, \citenamefont
		{Chumakov}, \citenamefont {Roth}, \citenamefont {Gonz{\'a}lez-Garc{\'\i}a},
		\citenamefont {Kraus},\ and\ \citenamefont
		{Schilling}}]{coupette2021percolation}%
	\BibitemOpen
	\bibfield  {author} {\bibinfo {author} {\bibfnamefont {F.}~\bibnamefont
			{Coupette}}, \bibinfo {author} {\bibfnamefont {L.}~\bibnamefont {Zhang}},
		\bibinfo {author} {\bibfnamefont {B.}~\bibnamefont {Kuttich}}, \bibinfo
		{author} {\bibfnamefont {A.}~\bibnamefont {Chumakov}}, \bibinfo {author}
		{\bibfnamefont {S.~V.}\ \bibnamefont {Roth}}, \bibinfo {author}
		{\bibfnamefont {L.}~\bibnamefont {Gonz{\'a}lez-Garc{\'\i}a}}, \bibinfo
		{author} {\bibfnamefont {T.}~\bibnamefont {Kraus}},\ and\ \bibinfo {author}
		{\bibfnamefont {T.}~\bibnamefont {Schilling}},\ }\bibfield  {journal}
	{\bibinfo  {journal} {J. Chem. Phys.}\ }\textbf {\bibinfo {volume} {155}},\
	\href {https://doi.org/10.1063/5.0058503} {10.1063/5.0058503} (\bibinfo
	{year} {2021})\BibitemShut {NoStop}%
	\bibitem [{\citenamefont {Fellay}\ \emph {et~al.}(2013)\citenamefont {Fellay},
		\citenamefont {Twist},\ and\ \citenamefont {Vanni}}]{fellay2013motion}%
	\BibitemOpen
	\bibfield  {author} {\bibinfo {author} {\bibfnamefont {L.~S.}\ \bibnamefont
			{Fellay}}, \bibinfo {author} {\bibfnamefont {C.}~\bibnamefont {Twist}},\ and\
		\bibinfo {author} {\bibfnamefont {M.}~\bibnamefont {Vanni}},\ }\href
	{https://doi.org/10.1007/s00707-013-0928-9} {\bibfield  {journal} {\bibinfo
			{journal} {Acta Mech.}\ }\textbf {\bibinfo {volume} {224}},\ \bibinfo {pages}
		{2225} (\bibinfo {year} {2013})}\BibitemShut {NoStop}%
	\bibitem [{\citenamefont {Trofa}\ and\ \citenamefont
		{D’Avino}(2020{\natexlab{a}})}]{trofa2020rheology}%
	\BibitemOpen
	\bibfield  {author} {\bibinfo {author} {\bibfnamefont {M.}~\bibnamefont
			{Trofa}}\ and\ \bibinfo {author} {\bibfnamefont {G.}~\bibnamefont
			{D’Avino}},\ }\href {https://doi.org/doi.org/10.3390/mi11040443} {\bibfield
		{journal} {\bibinfo  {journal} {Micromachines}\ }\textbf {\bibinfo {volume}
			{11}},\ \bibinfo {pages} {443} (\bibinfo {year}
		{2020}{\natexlab{a}})}\BibitemShut {NoStop}%
	\bibitem [{\citenamefont {Yu}\ and\ \citenamefont {Niu}(2024)}]{yu2024motion}%
	\BibitemOpen
	\bibfield  {author} {\bibinfo {author} {\bibfnamefont {Z.}~\bibnamefont
			{Yu}}\ and\ \bibinfo {author} {\bibfnamefont {X.}~\bibnamefont {Niu}},\
	}\bibfield  {journal} {\bibinfo  {journal} {Phys. Fluids}\ }\textbf {\bibinfo
		{volume} {36}},\ \href {https://doi.org/10.1063/5.0176759}
	{10.1063/5.0176759} (\bibinfo {year} {2024})\BibitemShut {NoStop}%
	\bibitem [{\citenamefont {Asylbekov}\ \emph {et~al.}(2021)\citenamefont
		{Asylbekov}, \citenamefont {Trunk}, \citenamefont {Krause},\ and\
		\citenamefont {Nirschl}}]{asylbekov2021microscale}%
	\BibitemOpen
	\bibfield  {author} {\bibinfo {author} {\bibfnamefont {E.}~\bibnamefont
			{Asylbekov}}, \bibinfo {author} {\bibfnamefont {R.}~\bibnamefont {Trunk}},
		\bibinfo {author} {\bibfnamefont {M.~J.}\ \bibnamefont {Krause}},\ and\
		\bibinfo {author} {\bibfnamefont {H.}~\bibnamefont {Nirschl}},\ }\href
	{https://doi.org/10.1002/ente.202000850} {\bibfield  {journal} {\bibinfo
			{journal} {Energy Technol.}\ }\textbf {\bibinfo {volume} {9}},\ \bibinfo
		{pages} {2000850} (\bibinfo {year} {2021})}\BibitemShut {NoStop}%
	\bibitem [{\citenamefont {Dizaji}\ \emph {et~al.}(2019)\citenamefont {Dizaji},
		\citenamefont {Marshall},\ and\ \citenamefont {Grant}}]{dizaji2019collision}%
	\BibitemOpen
	\bibfield  {author} {\bibinfo {author} {\bibfnamefont {F.~F.}\ \bibnamefont
			{Dizaji}}, \bibinfo {author} {\bibfnamefont {J.~S.}\ \bibnamefont
			{Marshall}},\ and\ \bibinfo {author} {\bibfnamefont {J.~R.}\ \bibnamefont
			{Grant}},\ }\href {https://doi.org/10.1017/jfm.2018.959} {\bibfield
		{journal} {\bibinfo  {journal} {J. Fluid Mech.}\ }\textbf {\bibinfo {volume}
			{862}},\ \bibinfo {pages} {592} (\bibinfo {year} {2019})}\BibitemShut
	{NoStop}%
	\bibitem [{\citenamefont {Trofa}\ and\ \citenamefont
		{D’Avino}(2020{\natexlab{b}})}]{trofa2020sedimentation}%
	\BibitemOpen
	\bibfield  {author} {\bibinfo {author} {\bibfnamefont {M.}~\bibnamefont
			{Trofa}}\ and\ \bibinfo {author} {\bibfnamefont {G.}~\bibnamefont
			{D’Avino}},\ }\href {https://doi.org/10.3390/app10093267} {\bibfield
		{journal} {\bibinfo  {journal} {Appl. Sci.}\ }\textbf {\bibinfo {volume}
			{10}},\ \bibinfo {pages} {3267} (\bibinfo {year}
		{2020}{\natexlab{b}})}\BibitemShut {NoStop}%
	\bibitem [{\citenamefont {Poon}\ and\ \citenamefont
		{Haw}(1997)}]{poon1997mesoscopic}%
	\BibitemOpen
	\bibfield  {author} {\bibinfo {author} {\bibfnamefont {W.~C.}\ \bibnamefont
			{Poon}}\ and\ \bibinfo {author} {\bibfnamefont {M.}~\bibnamefont {Haw}},\
	}\href {https://doi.org/10.1016/S0001-8686(97)90003-8} {\bibfield  {journal}
		{\bibinfo  {journal} {Adv. Colloid Interface Sci.}\ }\textbf {\bibinfo
			{volume} {73}},\ \bibinfo {pages} {71} (\bibinfo {year} {1997})}\BibitemShut
	{NoStop}%
	\bibitem [{\citenamefont {Evans}\ \emph {et~al.}(2008)\citenamefont {Evans},
		\citenamefont {Prasher}, \citenamefont {Fish}, \citenamefont {Meakin},
		\citenamefont {Phelan},\ and\ \citenamefont {Keblinski}}]{evans2008effect}%
	\BibitemOpen
	\bibfield  {author} {\bibinfo {author} {\bibfnamefont {W.}~\bibnamefont
			{Evans}}, \bibinfo {author} {\bibfnamefont {R.}~\bibnamefont {Prasher}},
		\bibinfo {author} {\bibfnamefont {J.}~\bibnamefont {Fish}}, \bibinfo {author}
		{\bibfnamefont {P.}~\bibnamefont {Meakin}}, \bibinfo {author} {\bibfnamefont
			{P.}~\bibnamefont {Phelan}},\ and\ \bibinfo {author} {\bibfnamefont
			{P.}~\bibnamefont {Keblinski}},\ }\href
	{https://doi.org/10.1016/j.ijheatmasstransfer.2007.10.017} {\bibfield
		{journal} {\bibinfo  {journal} {Int. J. Heat Mass Transfer}\ }\textbf
		{\bibinfo {volume} {51}},\ \bibinfo {pages} {1431} (\bibinfo {year}
		{2008})}\BibitemShut {NoStop}%
	\bibitem [{\citenamefont {Etheridge}\ \emph {et~al.}(2014)\citenamefont
		{Etheridge}, \citenamefont {Hurley}, \citenamefont {Zhang}, \citenamefont
		{Jeon}, \citenamefont {Ring}, \citenamefont {Hogan}, \citenamefont {Haynes},
		\citenamefont {Garwood},\ and\ \citenamefont
		{Bischof}}]{etheridge2014accounting}%
	\BibitemOpen
	\bibfield  {author} {\bibinfo {author} {\bibfnamefont {M.~L.}\ \bibnamefont
			{Etheridge}}, \bibinfo {author} {\bibfnamefont {K.~R.}\ \bibnamefont
			{Hurley}}, \bibinfo {author} {\bibfnamefont {J.}~\bibnamefont {Zhang}},
		\bibinfo {author} {\bibfnamefont {S.}~\bibnamefont {Jeon}}, \bibinfo {author}
		{\bibfnamefont {H.~L.}\ \bibnamefont {Ring}}, \bibinfo {author}
		{\bibfnamefont {C.}~\bibnamefont {Hogan}}, \bibinfo {author} {\bibfnamefont
			{C.~L.}\ \bibnamefont {Haynes}}, \bibinfo {author} {\bibfnamefont
			{M.}~\bibnamefont {Garwood}},\ and\ \bibinfo {author} {\bibfnamefont {J.~C.}\
			\bibnamefont {Bischof}},\ }\href {https://doi.org/10.1142/S2339547814500198}
	{\bibfield  {journal} {\bibinfo  {journal} {Technology}\ }\textbf {\bibinfo
			{volume} {2}},\ \bibinfo {pages} {214} (\bibinfo {year} {2014})}\BibitemShut
	{NoStop}%
	\bibitem [{\citenamefont {Katyal}\ \emph {et~al.}(2014)\citenamefont {Katyal},
		\citenamefont {Banerjee},\ and\ \citenamefont {Puri}}]{katyal2014fractal}%
	\BibitemOpen
	\bibfield  {author} {\bibinfo {author} {\bibfnamefont {N.}~\bibnamefont
			{Katyal}}, \bibinfo {author} {\bibfnamefont {V.}~\bibnamefont {Banerjee}},\
		and\ \bibinfo {author} {\bibfnamefont {S.}~\bibnamefont {Puri}},\ }\href
	{https://doi.org/10.1016/j.jqsrt.2014.01.017} {\bibfield  {journal} {\bibinfo
			{journal} {J. Quant. Spectrosc. Radiat. Transfer}\ }\textbf {\bibinfo
			{volume} {146}},\ \bibinfo {pages} {290} (\bibinfo {year}
		{2014})}\BibitemShut {NoStop}%
	\bibitem [{\citenamefont {Meakin}(1983)}]{meakin1983diffusion}%
	\BibitemOpen
	\bibfield  {author} {\bibinfo {author} {\bibfnamefont {P.}~\bibnamefont
			{Meakin}},\ }\href {https://doi.org/10.1103/PhysRevA.27.1495} {\bibfield
		{journal} {\bibinfo  {journal} {Phys. Rev. A}\ }\textbf {\bibinfo {volume}
			{27}},\ \bibinfo {pages} {1495} (\bibinfo {year} {1983})}\BibitemShut
	{NoStop}%
	\bibitem [{\citenamefont {Kl\"uppel}\ and\ \citenamefont
		{Heinrich}(1995)}]{kluppel1995fractal}%
	\BibitemOpen
	\bibfield  {author} {\bibinfo {author} {\bibfnamefont {M.}~\bibnamefont
			{Kl\"uppel}}\ and\ \bibinfo {author} {\bibfnamefont {G.}~\bibnamefont
			{Heinrich}},\ }\href {https://doi.org/10.5254/1.3538763} {\bibfield
		{journal} {\bibinfo  {journal} {Rubber Chem. Technol.}\ }\textbf {\bibinfo
			{volume} {68}},\ \bibinfo {pages} {623} (\bibinfo {year} {1995})}\BibitemShut
	{NoStop}%
	\bibitem [{\citenamefont {Zimmer}\ \emph {et~al.}(2024)\citenamefont {Zimmer},
		\citenamefont {Niebuur}, \citenamefont {Schaefer}, \citenamefont {Coupette},
		\citenamefont {T{\"a}nzel}, \citenamefont {Schilling},\ and\ \citenamefont
		{Kraus}}]{zimmer2024flow}%
	\BibitemOpen
	\bibfield  {author} {\bibinfo {author} {\bibfnamefont {B.}~\bibnamefont
			{Zimmer}}, \bibinfo {author} {\bibfnamefont {B.-J.}\ \bibnamefont {Niebuur}},
		\bibinfo {author} {\bibfnamefont {F.}~\bibnamefont {Schaefer}}, \bibinfo
		{author} {\bibfnamefont {F.}~\bibnamefont {Coupette}}, \bibinfo {author}
		{\bibfnamefont {V.}~\bibnamefont {T{\"a}nzel}}, \bibinfo {author}
		{\bibfnamefont {T.}~\bibnamefont {Schilling}},\ and\ \bibinfo {author}
		{\bibfnamefont {T.}~\bibnamefont {Kraus}},\ }\bibfield  {journal} {\bibinfo
		{journal} {arXiv preprint arXiv:2407.20318}\ }\href
	{https://doi.org/10.48550/arXiv.2407.20318} {10.48550/arXiv.2407.20318}
	(\bibinfo {year} {2024})\BibitemShut {NoStop}%
	\bibitem [{\citenamefont {Coupette}(2022)}]{coupette2022percolation}%
	\BibitemOpen
	\bibfield  {author} {\bibinfo {author} {\bibfnamefont {F.}~\bibnamefont
			{Coupette}},\ }\emph {\bibinfo {title} {Percolation: connecting the dots}},\
	\href@noop {} {Ph.D. thesis},\ \bibinfo  {school} {Dissertation,
		Universit{\"a}t Freiburg} (\bibinfo {year} {2022})\BibitemShut {NoStop}%
	\bibitem [{\citenamefont {Thompson}\ \emph {et~al.}(2022)\citenamefont
		{Thompson}, \citenamefont {Aktulga}, \citenamefont {Berger}, \citenamefont
		{Bolintineanu}, \citenamefont {Brown}, \citenamefont {Crozier}, \citenamefont
		{In't~Veld}, \citenamefont {Kohlmeyer}, \citenamefont {Moore}, \citenamefont
		{Nguyen} \emph {et~al.}}]{thompson2022lammps}%
	\BibitemOpen
	\bibfield  {author} {\bibinfo {author} {\bibfnamefont {A.~P.}\ \bibnamefont
			{Thompson}}, \bibinfo {author} {\bibfnamefont {H.~M.}\ \bibnamefont
			{Aktulga}}, \bibinfo {author} {\bibfnamefont {R.}~\bibnamefont {Berger}},
		\bibinfo {author} {\bibfnamefont {D.~S.}\ \bibnamefont {Bolintineanu}},
		\bibinfo {author} {\bibfnamefont {W.~M.}\ \bibnamefont {Brown}}, \bibinfo
		{author} {\bibfnamefont {P.~S.}\ \bibnamefont {Crozier}}, \bibinfo {author}
		{\bibfnamefont {P.~J.}\ \bibnamefont {In't~Veld}}, \bibinfo {author}
		{\bibfnamefont {A.}~\bibnamefont {Kohlmeyer}}, \bibinfo {author}
		{\bibfnamefont {S.~G.}\ \bibnamefont {Moore}}, \bibinfo {author}
		{\bibfnamefont {T.~D.}\ \bibnamefont {Nguyen}}, \emph {et~al.},\ }\href
	{https://doi.org/10.1016/j.cpc.2021.108171} {\bibfield  {journal} {\bibinfo
			{journal} {Comput. Phys. Commun.}\ }\textbf {\bibinfo {volume} {271}},\
		\bibinfo {pages} {108171} (\bibinfo {year} {2022})}\BibitemShut {NoStop}%
	\bibitem [{\citenamefont {Br{\"u}nger}\ \emph {et~al.}(1984)\citenamefont
		{Br{\"u}nger}, \citenamefont {Brooks~III},\ and\ \citenamefont
		{Karplus}}]{brunger1984stochastic}%
	\BibitemOpen
	\bibfield  {author} {\bibinfo {author} {\bibfnamefont {A.}~\bibnamefont
			{Br{\"u}nger}}, \bibinfo {author} {\bibfnamefont {C.~L.}\ \bibnamefont
			{Brooks~III}},\ and\ \bibinfo {author} {\bibfnamefont {M.}~\bibnamefont
			{Karplus}},\ }\href {https://doi.org/10.1016/0009-2614(84)80098-6} {\bibfield
		{journal} {\bibinfo  {journal} {Chem. Phys. Lett.}\ }\textbf {\bibinfo
			{volume} {105}},\ \bibinfo {pages} {495} (\bibinfo {year}
		{1984})}\BibitemShut {NoStop}%
	\bibitem [{\citenamefont {D{\"u}nweg}\ and\ \citenamefont
		{Paul}(1991)}]{dunweg1991brownian}%
	\BibitemOpen
	\bibfield  {author} {\bibinfo {author} {\bibfnamefont {B.}~\bibnamefont
			{D{\"u}nweg}}\ and\ \bibinfo {author} {\bibfnamefont {W.}~\bibnamefont
			{Paul}},\ }\href {https://doi.org/10.1142/S0129183191001037} {\bibfield
		{journal} {\bibinfo  {journal} {Int. J. Mod. Phys. C}\ }\textbf {\bibinfo
			{volume} {2}},\ \bibinfo {pages} {817} (\bibinfo {year} {1991})}\BibitemShut
	{NoStop}%
	\bibitem [{\citenamefont {Schroeder}\ \emph
		{et~al.}(2005{\natexlab{a}})\citenamefont {Schroeder}, \citenamefont
		{Teixeira}, \citenamefont {Shaqfeh},\ and\ \citenamefont
		{Chu}}]{schroeder2005dynamics}%
	\BibitemOpen
	\bibfield  {author} {\bibinfo {author} {\bibfnamefont {C.~M.}\ \bibnamefont
			{Schroeder}}, \bibinfo {author} {\bibfnamefont {R.~E.}\ \bibnamefont
			{Teixeira}}, \bibinfo {author} {\bibfnamefont {E.~S.}\ \bibnamefont
			{Shaqfeh}},\ and\ \bibinfo {author} {\bibfnamefont {S.}~\bibnamefont {Chu}},\
	}\href {https://doi.org/10.1021/ma0480796} {\bibfield  {journal} {\bibinfo
			{journal} {Macromolecules}\ }\textbf {\bibinfo {volume} {38}},\ \bibinfo
		{pages} {1967} (\bibinfo {year} {2005}{\natexlab{a}})}\BibitemShut {NoStop}%
	\bibitem [{\citenamefont {Jeffery}(1922)}]{jeffery1922motion}%
	\BibitemOpen
	\bibfield  {author} {\bibinfo {author} {\bibfnamefont {G.~B.}\ \bibnamefont
			{Jeffery}},\ }\href {https://doi.org/10.1098/rspa.1922.0078} {\bibfield
		{journal} {\bibinfo  {journal} {Proc. R. soc. Lond. Ser. A-Contain. Pap.
				Math. Phys. Character}\ }\textbf {\bibinfo {volume} {102}},\ \bibinfo {pages}
		{161} (\bibinfo {year} {1922})}\BibitemShut {NoStop}%
	\bibitem [{\citenamefont {Kuhnhold}\ and\ \citenamefont
		{Schilling}(2016)}]{kuhnhold2016isotropic}%
	\BibitemOpen
	\bibfield  {author} {\bibinfo {author} {\bibfnamefont {A.}~\bibnamefont
			{Kuhnhold}}\ and\ \bibinfo {author} {\bibfnamefont {T.}~\bibnamefont
			{Schilling}},\ }\bibfield  {journal} {\bibinfo  {journal} {The Journal of
			Chemical Physics}\ }\textbf {\bibinfo {volume} {145}},\ \href
	{https://doi.org/10.1063/1.4967718} {10.1063/1.4967718} (\bibinfo {year}
	{2016})\BibitemShut {NoStop}%
	\bibitem [{\citenamefont {Dhont}\ and\ \citenamefont
		{Briels}(2006)}]{dhont2006rod}%
	\BibitemOpen
	\bibfield  {author} {\bibinfo {author} {\bibfnamefont {J.}~\bibnamefont
			{Dhont}}\ and\ \bibinfo {author} {\bibfnamefont {W.~J.}\ \bibnamefont
			{Briels}},\ }\bibfield  {journal} {\bibinfo  {journal} {Soft Matter: Complex
			Colloidal Suspensions, edited by G. Gompper, M. Schick}\ }\textbf {\bibinfo
		{volume} {2}},\ \href {https://doi.org/10.1002/9783527617067}
	{10.1002/9783527617067} (\bibinfo {year} {2006})\BibitemShut {NoStop}%
	\bibitem [{\citenamefont {Aust}\ \emph {et~al.}(1999)\citenamefont {Aust},
		\citenamefont {Kr{\"o}ger},\ and\ \citenamefont {Hess}}]{aust1999structure}%
	\BibitemOpen
	\bibfield  {author} {\bibinfo {author} {\bibfnamefont {C.}~\bibnamefont
			{Aust}}, \bibinfo {author} {\bibfnamefont {M.}~\bibnamefont {Kr{\"o}ger}},\
		and\ \bibinfo {author} {\bibfnamefont {S.}~\bibnamefont {Hess}},\ }\href@noop
	{} {\bibfield  {journal} {\bibinfo  {journal} {Macromolecules}\ }\textbf
		{\bibinfo {volume} {32}},\ \bibinfo {pages} {5660} (\bibinfo {year}
		{1999})}\BibitemShut {NoStop}%
	\bibitem [{\citenamefont {Ripoll}\ \emph {et~al.}(2006)\citenamefont {Ripoll},
		\citenamefont {Winkler},\ and\ \citenamefont {Gompper}}]{ripoll2006star}%
	\BibitemOpen
	\bibfield  {author} {\bibinfo {author} {\bibfnamefont {M.}~\bibnamefont
			{Ripoll}}, \bibinfo {author} {\bibfnamefont {R.}~\bibnamefont {Winkler}},\
		and\ \bibinfo {author} {\bibfnamefont {G.}~\bibnamefont {Gompper}},\ }\href
	{https://doi.org/10.1103/PhysRevLett.96.188302} {\bibfield  {journal}
		{\bibinfo  {journal} {Phys. Rev. Lett.}\ }\textbf {\bibinfo {volume} {96}},\
		\bibinfo {pages} {188302} (\bibinfo {year} {2006})}\BibitemShut {NoStop}%
	\bibitem [{\citenamefont {Winkler}\ \emph {et~al.}(2004)\citenamefont
		{Winkler}, \citenamefont {Mussawisade}, \citenamefont {Ripoll},\ and\
		\citenamefont {Gompper}}]{winkler2004rod}%
	\BibitemOpen
	\bibfield  {author} {\bibinfo {author} {\bibfnamefont {R.}~\bibnamefont
			{Winkler}}, \bibinfo {author} {\bibfnamefont {K.}~\bibnamefont
			{Mussawisade}}, \bibinfo {author} {\bibfnamefont {M.}~\bibnamefont
			{Ripoll}},\ and\ \bibinfo {author} {\bibfnamefont {G.}~\bibnamefont
			{Gompper}},\ }\href {https://doi.org/10.1088/0953-8984/16/38/012} {\bibfield
		{journal} {\bibinfo  {journal} {J. Phys.: Condens. Matter}\ }\textbf
		{\bibinfo {volume} {16}},\ \bibinfo {pages} {S3941} (\bibinfo {year}
		{2004})}\BibitemShut {NoStop}%
	\bibitem [{\citenamefont {Bretherton}(1962)}]{bretherton1962motion}%
	\BibitemOpen
	\bibfield  {author} {\bibinfo {author} {\bibfnamefont {F.~P.}\ \bibnamefont
			{Bretherton}},\ }\href {https://doi.org/10.1017/S002211206200124X} {\bibfield
		{journal} {\bibinfo  {journal} {J. Fluid Mech.}\ }\textbf {\bibinfo {volume}
			{14}},\ \bibinfo {pages} {284} (\bibinfo {year} {1962})}\BibitemShut
	{NoStop}%
	\bibitem [{\citenamefont {Hinch}\ and\ \citenamefont
		{Leal}(1979)}]{hinch1979rotation}%
	\BibitemOpen
	\bibfield  {author} {\bibinfo {author} {\bibfnamefont {E.}~\bibnamefont
			{Hinch}}\ and\ \bibinfo {author} {\bibfnamefont {L.}~\bibnamefont {Leal}},\
	}\href {https://doi.org/10.1017/S002211207900077X} {\bibfield  {journal}
		{\bibinfo  {journal} {J. Fluid Mech.}\ }\textbf {\bibinfo {volume} {92}},\
		\bibinfo {pages} {591} (\bibinfo {year} {1979})}\BibitemShut {NoStop}%
	\bibitem [{\citenamefont {Xu}\ \emph {et~al.}(2014)\citenamefont {Xu},
		\citenamefont {Chen},\ and\ \citenamefont {An}}]{xu2014shear}%
	\BibitemOpen
	\bibfield  {author} {\bibinfo {author} {\bibfnamefont {X.}~\bibnamefont
			{Xu}}, \bibinfo {author} {\bibfnamefont {J.}~\bibnamefont {Chen}},\ and\
		\bibinfo {author} {\bibfnamefont {L.}~\bibnamefont {An}},\ }\bibfield
	{journal} {\bibinfo  {journal} {J. Chem. Phys.}\ }\textbf {\bibinfo {volume}
		{140}},\ \href {https://doi.org/doi.org/10.1063/1.4873709}
	{doi.org/10.1063/1.4873709} (\bibinfo {year} {2014})\BibitemShut {NoStop}%
	\bibitem [{\citenamefont {Chen}\ \emph {et~al.}(2013)\citenamefont {Chen},
		\citenamefont {Chen},\ and\ \citenamefont {An}}]{chen2013tumbling}%
	\BibitemOpen
	\bibfield  {author} {\bibinfo {author} {\bibfnamefont {W.}~\bibnamefont
			{Chen}}, \bibinfo {author} {\bibfnamefont {J.}~\bibnamefont {Chen}},\ and\
		\bibinfo {author} {\bibfnamefont {L.}~\bibnamefont {An}},\ }\href
	{https://doi.org/10.1039/C3SM50352F} {\bibfield  {journal} {\bibinfo
			{journal} {Soft Matter}\ }\textbf {\bibinfo {volume} {9}},\ \bibinfo {pages}
		{4312} (\bibinfo {year} {2013})}\BibitemShut {NoStop}%
	\bibitem [{\citenamefont {Sabli{\'c}}\ \emph {et~al.}(2017)\citenamefont
		{Sabli{\'c}}, \citenamefont {Praprotnik},\ and\ \citenamefont
		{Delgado-Buscalioni}}]{sablic2017deciphering}%
	\BibitemOpen
	\bibfield  {author} {\bibinfo {author} {\bibfnamefont {J.}~\bibnamefont
			{Sabli{\'c}}}, \bibinfo {author} {\bibfnamefont {M.}~\bibnamefont
			{Praprotnik}},\ and\ \bibinfo {author} {\bibfnamefont {R.}~\bibnamefont
			{Delgado-Buscalioni}},\ }\href {https://doi.org/10.1039/C7SM00364A}
	{\bibfield  {journal} {\bibinfo  {journal} {Soft Matter}\ }\textbf {\bibinfo
			{volume} {13}},\ \bibinfo {pages} {4971} (\bibinfo {year}
		{2017})}\BibitemShut {NoStop}%
	\bibitem [{\citenamefont {Huang}\ \emph {et~al.}(2011)\citenamefont {Huang},
		\citenamefont {Sutmann}, \citenamefont {Gompper},\ and\ \citenamefont
		{Winkler}}]{huang2011tumbling}%
	\BibitemOpen
	\bibfield  {author} {\bibinfo {author} {\bibfnamefont {C.-C.}\ \bibnamefont
			{Huang}}, \bibinfo {author} {\bibfnamefont {G.}~\bibnamefont {Sutmann}},
		\bibinfo {author} {\bibfnamefont {G.}~\bibnamefont {Gompper}},\ and\ \bibinfo
		{author} {\bibfnamefont {R.}~\bibnamefont {Winkler}},\ }\href
	{https://doi.org/10.1209/0295-5075/93/54004} {\bibfield  {journal} {\bibinfo
			{journal} {Europhys. Lett.}\ }\textbf {\bibinfo {volume} {93}},\ \bibinfo
		{pages} {54004} (\bibinfo {year} {2011})}\BibitemShut {NoStop}%
	\bibitem [{\citenamefont {Peng}\ \emph {et~al.}(2024)\citenamefont {Peng},
		\citenamefont {Liu},\ and\ \citenamefont {Wang}}]{peng2024unveiling}%
	\BibitemOpen
	\bibfield  {author} {\bibinfo {author} {\bibfnamefont {B.}~\bibnamefont
			{Peng}}, \bibinfo {author} {\bibfnamefont {L.}~\bibnamefont {Liu}},\ and\
		\bibinfo {author} {\bibfnamefont {D.}~\bibnamefont {Wang}},\ }\bibfield
	{journal} {\bibinfo  {journal} {J. Chem. Phys.}\ }\textbf {\bibinfo {volume}
		{160}},\ \href {https://doi.org/10.1063/5.0198272} {10.1063/5.0198272}
	(\bibinfo {year} {2024})\BibitemShut {NoStop}%
	\bibitem [{\citenamefont {Nikoubashman}\ and\ \citenamefont
		{Howard}(2017)}]{nikoubashman2017equilibrium}%
	\BibitemOpen
	\bibfield  {author} {\bibinfo {author} {\bibfnamefont {A.}~\bibnamefont
			{Nikoubashman}}\ and\ \bibinfo {author} {\bibfnamefont {M.~P.}\ \bibnamefont
			{Howard}},\ }\href {https://doi.org/10.1021/acs.macromol.7b01876} {\bibfield
		{journal} {\bibinfo  {journal} {Macromolecules}\ }\textbf {\bibinfo {volume}
			{50}},\ \bibinfo {pages} {8279} (\bibinfo {year} {2017})}\BibitemShut
	{NoStop}%
	\bibitem [{\citenamefont {Winkler}(2006)}]{winkler2006semiflexible}%
	\BibitemOpen
	\bibfield  {author} {\bibinfo {author} {\bibfnamefont {R.~G.}\ \bibnamefont
			{Winkler}},\ }\href {https://doi.org/10.1103/PhysRevLett.97.128301}
	{\bibfield  {journal} {\bibinfo  {journal} {Phys. Rev. Lett.}\ }\textbf
		{\bibinfo {volume} {97}},\ \bibinfo {pages} {128301} (\bibinfo {year}
		{2006})}\BibitemShut {NoStop}%
	\bibitem [{\citenamefont {Li}\ \emph {et~al.}(2021)\citenamefont {Li},
		\citenamefont {Gompper},\ and\ \citenamefont {Ripoll}}]{li2021tumbling}%
	\BibitemOpen
	\bibfield  {author} {\bibinfo {author} {\bibfnamefont {R.}~\bibnamefont
			{Li}}, \bibinfo {author} {\bibfnamefont {G.}~\bibnamefont {Gompper}},\ and\
		\bibinfo {author} {\bibfnamefont {M.}~\bibnamefont {Ripoll}},\ }\href
	{https://doi.org/10.1021/acs.macromol.0c01651} {\bibfield  {journal}
		{\bibinfo  {journal} {Macromolecules}\ }\textbf {\bibinfo {volume} {54}},\
		\bibinfo {pages} {812} (\bibinfo {year} {2021})}\BibitemShut {NoStop}%
	\bibitem [{\citenamefont {Schroeder}\ \emph
		{et~al.}(2005{\natexlab{b}})\citenamefont {Schroeder}, \citenamefont
		{Teixeira}, \citenamefont {Shaqfeh},\ and\ \citenamefont
		{Chu}}]{schroeder2005characteristic}%
	\BibitemOpen
	\bibfield  {author} {\bibinfo {author} {\bibfnamefont {C.~M.}\ \bibnamefont
			{Schroeder}}, \bibinfo {author} {\bibfnamefont {R.~E.}\ \bibnamefont
			{Teixeira}}, \bibinfo {author} {\bibfnamefont {E.~S.}\ \bibnamefont
			{Shaqfeh}},\ and\ \bibinfo {author} {\bibfnamefont {S.}~\bibnamefont {Chu}},\
	}\href {https://doi.org/10.1103/PhysRevLett.95.018301} {\bibfield  {journal}
		{\bibinfo  {journal} {Phys. Rev. Lett.}\ }\textbf {\bibinfo {volume} {95}},\
		\bibinfo {pages} {018301} (\bibinfo {year} {2005}{\natexlab{b}})}\BibitemShut
	{NoStop}%
	\bibitem [{\citenamefont {Lang}\ \emph {et~al.}(2014)\citenamefont {Lang},
		\citenamefont {Obermayer},\ and\ \citenamefont {Frey}}]{lang2014dynamics}%
	\BibitemOpen
	\bibfield  {author} {\bibinfo {author} {\bibfnamefont {P.~S.}\ \bibnamefont
			{Lang}}, \bibinfo {author} {\bibfnamefont {B.}~\bibnamefont {Obermayer}},\
		and\ \bibinfo {author} {\bibfnamefont {E.}~\bibnamefont {Frey}},\ }\href
	{https://doi.org/10.1103/PhysRevE.89.022606} {\bibfield  {journal} {\bibinfo
			{journal} {Phys. Rev. E}\ }\textbf {\bibinfo {volume} {89}},\ \bibinfo
		{pages} {022606} (\bibinfo {year} {2014})}\BibitemShut {NoStop}%
	\bibitem [{\citenamefont {Saha~Dalal}\ \emph {et~al.}(2012)\citenamefont
		{Saha~Dalal}, \citenamefont {Albaugh}, \citenamefont {Hoda},\ and\
		\citenamefont {Larson}}]{saha2012tumbling}%
	\BibitemOpen
	\bibfield  {author} {\bibinfo {author} {\bibfnamefont {I.}~\bibnamefont
			{Saha~Dalal}}, \bibinfo {author} {\bibfnamefont {A.}~\bibnamefont {Albaugh}},
		\bibinfo {author} {\bibfnamefont {N.}~\bibnamefont {Hoda}},\ and\ \bibinfo
		{author} {\bibfnamefont {R.~G.}\ \bibnamefont {Larson}},\ }\href
	{https://doi.org/10.1021/ma3014349} {\bibfield  {journal} {\bibinfo
			{journal} {Macromolecules}\ }\textbf {\bibinfo {volume} {45}},\ \bibinfo
		{pages} {9493} (\bibinfo {year} {2012})}\BibitemShut {NoStop}%
	\bibitem [{\citenamefont {Winkler}\ \emph {et~al.}(2014)\citenamefont
		{Winkler}, \citenamefont {Fedosov},\ and\ \citenamefont
		{Gompper}}]{winkler2014dynamical}%
	\BibitemOpen
	\bibfield  {author} {\bibinfo {author} {\bibfnamefont {R.~G.}\ \bibnamefont
			{Winkler}}, \bibinfo {author} {\bibfnamefont {D.~A.}\ \bibnamefont
			{Fedosov}},\ and\ \bibinfo {author} {\bibfnamefont {G.}~\bibnamefont
			{Gompper}},\ }\href {https://doi.org//10.1016/j.cocis.2014.09.005} {\bibfield
		{journal} {\bibinfo  {journal} {Curr. Opin. Colloid Interface Sci.}\
		}\textbf {\bibinfo {volume} {19}},\ \bibinfo {pages} {594} (\bibinfo {year}
		{2014})}\BibitemShut {NoStop}%
	\bibitem [{\citenamefont {Zwanzig}(1965)}]{zwanzig1965time}%
	\BibitemOpen
	\bibfield  {author} {\bibinfo {author} {\bibfnamefont {R.}~\bibnamefont
			{Zwanzig}},\ }\href {https://doi.org/10.1146/annurev.pc.16.100165.000435}
	{\bibfield  {journal} {\bibinfo  {journal} {Annu. Rev. Phys. Chem.}\ }\textbf
		{\bibinfo {volume} {16}},\ \bibinfo {pages} {67} (\bibinfo {year}
		{1965})}\BibitemShut {NoStop}%
	\bibitem [{\citenamefont {Vermant}\ and\ \citenamefont
		{Solomon}(2005)}]{vermant2005flow}%
	\BibitemOpen
	\bibfield  {author} {\bibinfo {author} {\bibfnamefont {J.}~\bibnamefont
			{Vermant}}\ and\ \bibinfo {author} {\bibfnamefont {M.~J.}\ \bibnamefont
			{Solomon}},\ }\href {https://doi.org/10.1088/0953-8984/17/4/R02} {\bibfield
		{journal} {\bibinfo  {journal} {J. Phys.: Condens. Matter}\ }\textbf
		{\bibinfo {volume} {17}},\ \bibinfo {pages} {R187} (\bibinfo {year}
		{2005})}\BibitemShut {NoStop}%
	\bibitem [{\citenamefont {Richards}\ \emph {et~al.}(2023)\citenamefont
		{Richards}, \citenamefont {Ramos},\ and\ \citenamefont
		{Liu}}]{richards2023review}%
	\BibitemOpen
	\bibfield  {author} {\bibinfo {author} {\bibfnamefont {J.~J.}\ \bibnamefont
			{Richards}}, \bibinfo {author} {\bibfnamefont {P.~Z.}\ \bibnamefont
			{Ramos}},\ and\ \bibinfo {author} {\bibfnamefont {Q.}~\bibnamefont {Liu}},\
	}\href {https://doi.org/10.3389/fphy.2023.1245847} {\bibfield  {journal}
		{\bibinfo  {journal} {Frontiers in Physics}\ }\textbf {\bibinfo {volume}
			{11}},\ \bibinfo {pages} {1245847} (\bibinfo {year} {2023})}\BibitemShut
	{NoStop}%
	\bibitem [{\citenamefont {Brady}(2001)}]{brady2001computer}%
	\BibitemOpen
	\bibfield  {author} {\bibinfo {author} {\bibfnamefont {J.~F.}\ \bibnamefont
			{Brady}},\ }\href {https://doi.org/10.1016/S0009-2509(00)00475-9} {\bibfield
		{journal} {\bibinfo  {journal} {Chem. Eng. Sci.}\ }\textbf {\bibinfo {volume}
			{56}},\ \bibinfo {pages} {2921} (\bibinfo {year} {2001})}\BibitemShut
	{NoStop}%
	\bibitem [{\citenamefont {Varga}\ \emph {et~al.}(2019)\citenamefont {Varga},
		\citenamefont {Grenard}, \citenamefont {Pecorario}, \citenamefont {Taberlet},
		\citenamefont {Dolique}, \citenamefont {Manneville}, \citenamefont {Divoux},
		\citenamefont {McKinley},\ and\ \citenamefont
		{Swan}}]{varga2019hydrodynamics}%
	\BibitemOpen
	\bibfield  {author} {\bibinfo {author} {\bibfnamefont {Z.}~\bibnamefont
			{Varga}}, \bibinfo {author} {\bibfnamefont {V.}~\bibnamefont {Grenard}},
		\bibinfo {author} {\bibfnamefont {S.}~\bibnamefont {Pecorario}}, \bibinfo
		{author} {\bibfnamefont {N.}~\bibnamefont {Taberlet}}, \bibinfo {author}
		{\bibfnamefont {V.}~\bibnamefont {Dolique}}, \bibinfo {author} {\bibfnamefont
			{S.}~\bibnamefont {Manneville}}, \bibinfo {author} {\bibfnamefont
			{T.}~\bibnamefont {Divoux}}, \bibinfo {author} {\bibfnamefont {G.~H.}\
			\bibnamefont {McKinley}},\ and\ \bibinfo {author} {\bibfnamefont {J.~W.}\
			\bibnamefont {Swan}},\ }\href {https://doi.org/10.1073/pnas.1901370116}
	{\bibfield  {journal} {\bibinfo  {journal} {Proc. Natl. Acad. Sci. U.S.A.}\
		}\textbf {\bibinfo {volume} {116}},\ \bibinfo {pages} {12193} (\bibinfo
		{year} {2019})}\BibitemShut {NoStop}%
	\bibitem [{\citenamefont {Nagel}(2022)}]{prettypyplot}%
	\BibitemOpen
	\bibfield  {author} {\bibinfo {author} {\bibfnamefont {D.}~\bibnamefont
			{Nagel}},\ }\href@noop {} {\bibinfo {title} {Prettypyplot: publication ready
			matplotlib figures made simple}} (\bibinfo {year} {2022}),\ \bibinfo {note}
	{{Z}enodo: 10.5281/zenodo.7278312}\BibitemShut {NoStop}%
\end{thebibliography}
\end{document}